\documentclass[a4paper,fleqn,usenatbib]{mnras}
\usepackage{graphicx}	
\usepackage{color}
\usepackage{soul}
\usepackage{bm}
\usepackage{amsmath,amssymb}
\usepackage{enumitem} 

\definecolor{ultramarine}{rgb}{0.07, 0.04, 0.56}
\newcommand{\kms}{\,km\,s$^{-1}$}
\newcommand{\kmskpc}{\,km\,s$^{-1}$\,kpc$^{-1}$}

\title[Double-barred Galaxy NGC\,3504]{Morphological and Kinematical Analysis of the Double-barred Galaxy NGC\,3504 Using ALMA CO (2--1) Data}

\author[Wu et al.]{
Yu-Ting Wu,$^{1}$\thanks{E-mail: yuting.wu@nao.ac.jp}
Alfonso Trejo,$^{2}$
Daniel Espada$^{3,4}$
and Yusuke Miyamoto$^{1}$
\\
$^{1}$National Astronomical Observatory of Japan, Mitaka, Tokyo 181-8588, Japan\\
$^{2}$Institute of Astronomy and Astrophysics, Academia Sinica, Taipei 10617, Taiwan\\
$^{3}$SKA Organisation, Lower Withington, Macclesfield, Cheshire SK11 9DL, UK\\
$^{4}$Departamento de F\'isica Te\'orica y del Cosmos, Campus de Fuentenueva, Universidad de Granada, E18071, Granada, Spain
}

\date{Accepted 2021 April 13. Received 2021 April 12; in original form 2020 July 5}

\pubyear{2021}

\begin{document}
\label{firstpage}
\pagerange{\pageref{firstpage}--\pageref{lastpage}}
\maketitle

\begin{abstract}
We present results obtained from ALMA CO (2--1) data of the double-barred galaxy NGC\,3504. 
With three times higher angular resolution ($\sim$ 0\farcs8) than previous studies, our observations reveal an inner molecular gas bar, a nuclear ring, and four inner spiral arm-like structures in the central 1\,kpc region. Furthermore, 
the CO emission is clearly aligned with the two dust lanes in the outer bar region, with differences in shape and intensity between them. 
The total molecular gas mass in the observed region ($50\arcsec \times 57\arcsec$) is estimated to be $\sim 3. 1\times 10^9 \, {\rm M}_{\odot}$, which is 17 per cent of the stellar mass.
We used the Kinemetry package to fit the velocity field and found that circular motion strongly dominates at $R= 0.3-0.8$\,kpc, but radial motion becomes important at $R<0.3$\,kpc and $R=1.0-2.5$\,kpc, which is expected due to the presence of the inner and outer bars. 
Finally, assuming that the gas moves along the dust lanes in the bar rotating frame, we derived the pattern speed of the outer bar to be $ 18\pm5$\kmskpc, the average streaming velocities on each of the two dust lanes to be 165 and 221\kms, and the total mass inflow rate along the dust lanes to be 12\,{\rm M}$_{\odot}$\,yr$^{-1}$. Our results give a new example of an inner gas bar within a gas-rich double-barred galaxy and suggest that the formation of  double-barred galaxies could be associated with the existence of such gas structures.
\end{abstract}



\section{Introduction}
Both external processes, such as galaxy collisions and mergers, and internal processes in galaxies, such as the actions of bars, play important roles in the evolution of galaxies. For example, galaxy collisions can not only change the shape of galaxies but also enhance their star formation rate \citep{Toomre72,Teyssier10}. On the other hand, regarding the internal processes, many studies have been conducted to understand how non-axisymmetric potentials, such as bars or spirals, redistribute the angular momentum within galaxies and then change the distribution of stars and gas. For example, the gas inflow driven by bars may lead to gas concentration near the galactic center, such as the Central Molecular Zone in our Galaxy, and change the star formation rate \citep{Schwarz81, Athanassoula92, Kruijssen14}. 
In such case, knowing the mass inflow rate of gas in barred galaxies is essential \citep{Regan97, Sormani19}.

On large scales, bars, spiral arms or dynamical perturbations, such as galaxy collisions, mergers and other minor accretions, can effectively drive gas inflow down to kpc scales  \citep{Heckman86, Combes88,  Hernquist89, Barnes92}. However, gas may settle into a nuclear ring, which is usually associated with resonance phenomena, and cannot easily reach the nucleus of a galaxy \citep{Patsis00}. To further reduce the angular momentum and allow the inflow of gas to the galactic center ($\sim100$ pc
scale), other mechanisms are needed, such as the existence of an inner bar or dissipation in a warped nuclear disk \citep{Shlosman89, Pringle96, Schinnerer00, Shu90}. 

As mentioned in \cite{Shlosman89}, the importance of the inner bar in double-barred galaxies is that the inner bar inside the inner Lindblad resonance of the outer bar could possibly drive gas inflow to fuel active galactic nuclei. However, the formation mechanism of the inner bar is still unclear.
Different numerical simulations have been conducted to understand the formation of double-barred galaxies and can be classified into two major categories: either with or without gas. In the first case, the well-known study by \cite{Friedli93} showed that a gaseous inner bar can be formed either simultaneously or a few 100 Myr after the outer bar, and then decouple from the outer bar with a different pattern speed. 
Note that the required gas mass in this scenario is about 10 per cent of the stellar mass of the galaxy in their simulations.
On the other hand, in the second case, pure collisionless simulations demonstrate the feasibility of forming double-barred galaxies under particular conditions, such as with a rapidly rotating pseudo-bulge (which can also be referred to as circumnuclear stellar disk) \citep{Shen09} or with a dynamically cooled inner disk embedded in a hotter outer disk \citep{Du15}. Therefore, the first key step to distinguish these two competing scenarios is to resolve the gaseous inner bar using high-resolution observations and determine the gas mass in double-barred galaxies.
In addition, studying the pattern speed of the outer bar can be used to examine the location of the inner Lindblad resonance and compare it with the locations of the nuclear ring and the inner bar.

\begin{figure*}
\includegraphics[scale=0.55, clip=true, trim = 0 0 0 0]{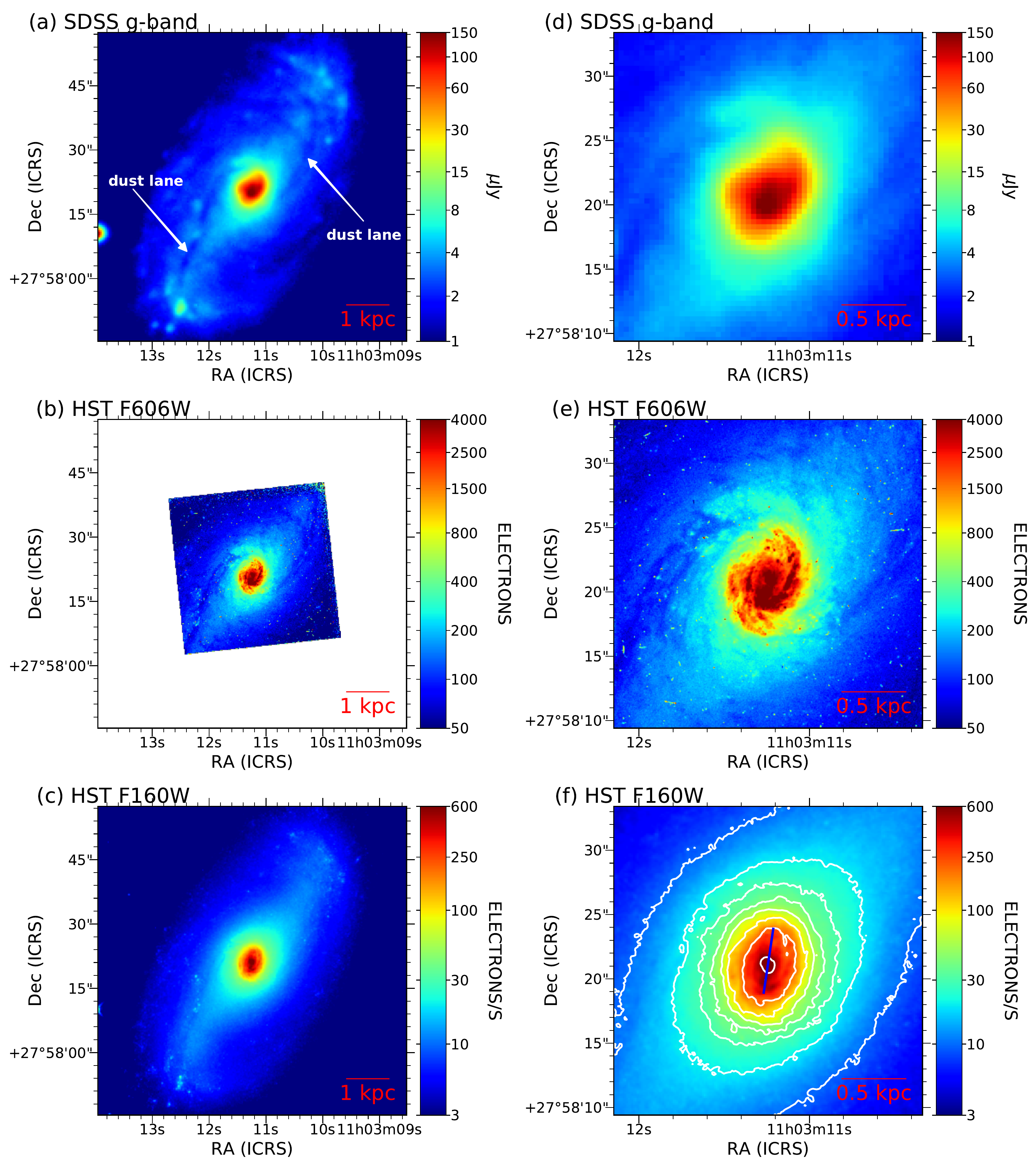}
 \caption{(a) to (c): The SDSS $g$-band, HST/WFPC2 F606W, and HST/WFC3 F160W images of NGC\,3504. The white arrows indicate the dust lanes seen in the SDSS $g$-band image. (d) to (f) are the same as panels (a) to (c), but showing the central 2.4 kpc region. The contour levels in panel (f) are 10, 20, 30, 40, 60, 100, 200, and 600 electrons/s. The blue line represents the major axis of the inner bar with a length of 5\farcs2 and position angle of 172\degr.}
\label {fig:NGC3504}
\end{figure*}

Our target NGC\,3504 has been classified as an early type barred spiral galaxy, (R)SAB(s)ab \citep{Vaucouleurs91}, as well as a potential double-barred galaxy.
Although the isophotes on the near-infrared (NIR) image show clear evidence for an inner stellar bar, this has also been interpreted as a possible secondary bright source near the galactic nucleus \citep{Perez00,Erwin04}.
In addition, NGC\,3504 has been classified as a "circumnuclear starburst", with a higher star formation rate and star formation rate per mass unit of molecular hydrogen than most other barred galaxies reported in \cite{Jogee2005J}.
Fig.~\ref{fig:NGC3504} shows the SDSS $g$-band, HST/WFPC2 F606W, and HST/WFC3 F160W images of NGC\,3504. 
The two dust lanes in the outer bar can be seen more clearly in the SDSS $g$-band image, and are indicated with two white arrows in Fig.~\ref{fig:NGC3504} (a). 
Also, with the higher angular resolution of the HST F606W optical image shown in Fig.~\ref{fig:NGC3504} (e), we see some nuclear spiral structures and dust lanes in the circumnuclear region.
Furthermore, Fig.~\ref{fig:NGC3504} (f) shows the inner stellar bar in the HST F160W NIR image.
The contours in Fig.~\ref{fig:NGC3504} (f) are used to emphasize the change of the position angle from the inner bar to the outer bar. The blue line denotes the major axis of the inner bar with a length of 5\farcs2 and position angle of 172\degr. The size ($L_{\rm bar}$) and the position angle were estimated using the ellipticity and position angle curves as functions of radius on the $K$-band image \citep{Perez00} by following the procedure of \cite{Erwin04}.

CO line observations of NGC\,3504 have been obtained with single-dish  \citep{Young84, planesas97, kuno00} and interferometric \citep{kenney93} telescopes. With the best resolution of $2\farcs5$, achieved at that time by OVRO,  \cite{kenney93} found that the CO distribution in the galactic central region is nearly azimuthally symmetric and can be described by an exponential profile. 
Furthermore, by assuming the corotation radius is located at the end of the outer bar, they derived the pattern speed of the  outer bar to be 77\kmskpc\ and the two inner Lindblad resonances at $r\sim 2\arcsec$ and $5\arcsec$--$12\arcsec$ (0.2 and $0.5-1.2$\,kpc, respectively). 
On the other hand, \cite{kuno00} derived an upper limit of the pattern speed of 41\kmskpc\ by assuming that the gas on the dust lanes  in the outer bar mainly flows along the dust lanes. This assumption implies that the velocity component perpendicular to the outer bar is much smaller than the component parallel to the bar and thus allows to obtain an upper limit 
of the pattern speed of the outer bar. If this upper limit is close to the pattern speed of the outer bar, \cite{kuno00} suggested the existence of two inner Lindblad resonances  located at $r\sim 1\arcsec$ and $16\arcsec$ (0.1 and 1.6\,kpc, respectively).

Since NGC\,3504 is rich in molecular gas emission and classified as a double-barred galaxy, this galaxy is unique to investigate the properties of molecular gas in 
its type. In this paper, we present high resolution CO (2--1) data and study its morphology and kinematics. Our paper is organized as follows. In Section 2, we present our ALMA CO (2--1) observations and adopted calibration. The CO channel maps, distribution, and velocity field are presented in Section 3. In Section 4, we describe the kinematic modelling and the fitting results, including the basic galactic parameters, circular and non-circular velocity fields. 
We calculate the gas streaming velocity along the dust lanes and the pattern speed of the outer bar in Sections 5 and 6. Finally, we present our discussion and conclusions in Sections 7 and 8. 

\section{Observations and Data Reduction}
\begin{figure}
\includegraphics[scale=0.46, clip=true, trim = 0 0 0 0]{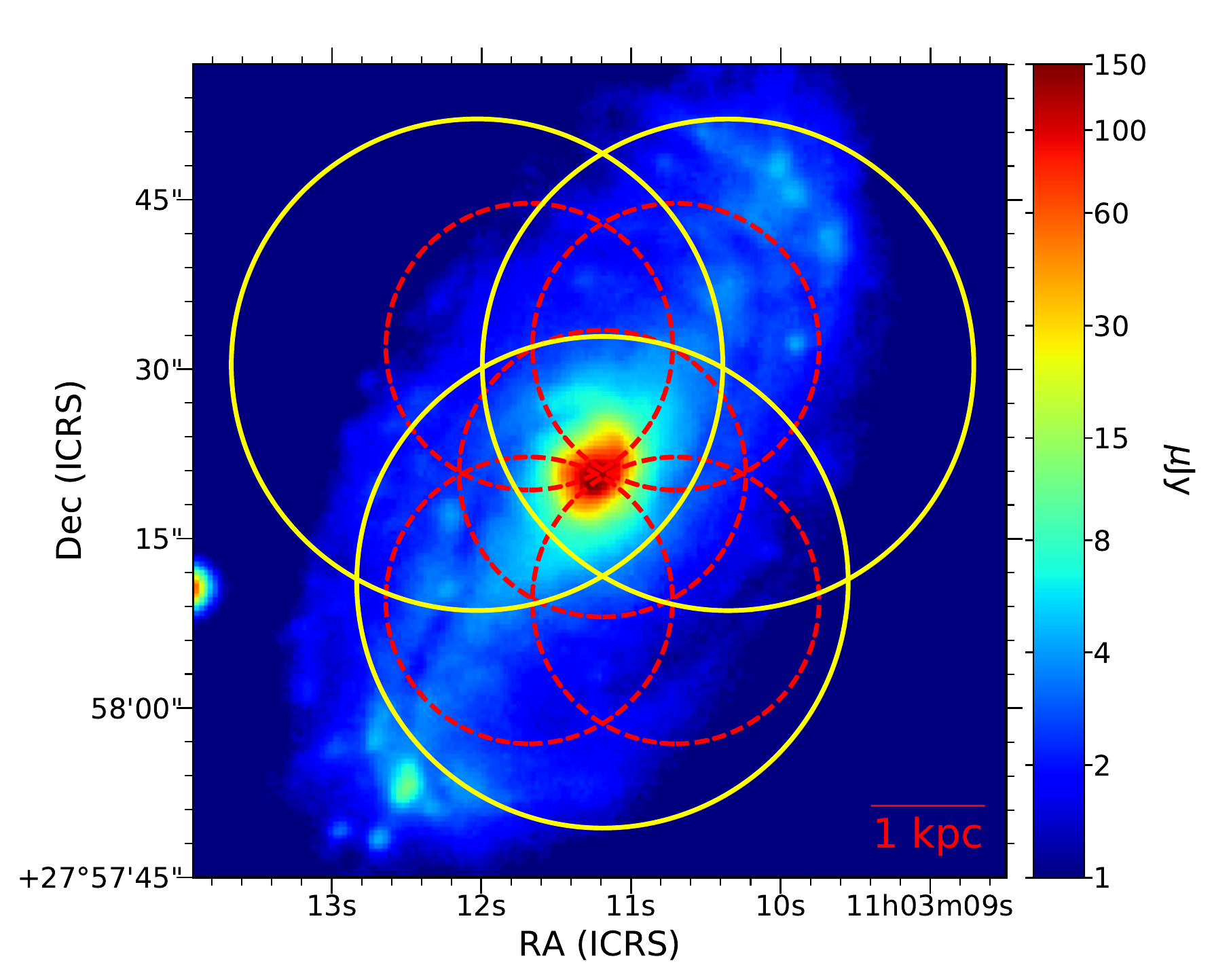}
 \caption{Observed ALMA pointings overlaid on the SDSS $g$-band image of NGC\,3504. The red and yellow circles show the five pointings of each 12m array and the three pointings of the 7m array, respectively. The size of the red and yellow circles are the primary beams of the 12m and 7m antennas, 25\farcs4 and 43\farcs5. 
 }
\label {fig:mosaic_coverage}
\end{figure}

Our ALMA CO (2--1) observations presented here
(project 2016.1.00650.S) were carried out on 21 November 2016 and 18 March 2017 with the two 12m array configurations C43-4 and C43-1, and on 12 and 17 November 2016 with the ACA 7m array.
Table~\ref{table:ObserPara} shows the summary of the observational parameters.
The mosaic area was covered with five pointings for each of the two 12m arrays and three pointings for the 7m array, as shown in Fig.~\ref{fig:mosaic_coverage}.
Note that the International Celestial Reference System (ICRS) is adopted in this paper.
The total on-source time for the C43-4, C43-1, and 7m array observations were about 34, 10, and 100 minutes, respectively. 
The four spectral windows were centered at the rest frequencies 230.538, 231.901, 244.936, and 246.819\,GHz with bandwidths
of 1.875\,GHz and spectral resolutions of 5.10, 5.07, 4.80, and 38.15\kms. 
The primary beams of the 12m and 7m antennas are 25\farcs4 and 43\farcs5 at the observed frequency of 229.367\,GHz, which is used to observe the CO (2--1) line. 
Furthermore, in our combined data, given the 5th percentile of the uv-distance $L_5\approx 9$\,m, the maximum recoverable scale per pointing is $\theta_{MRS} \approx \frac{0.983\lambda}{L_5}\approx30\arcsec$, where $\lambda$ is the observed wavelength.

\begin{table*}   
\centering
\caption{ALMA Observational Parameters}
 \begin{tabular}{lccccc}
 \hline
 \hline
 \multicolumn{2}{l}{Parameter}& \multicolumn{2}{c}{12m Array} & \multicolumn{2}{c}{ACA (Morita Array)}\\
 \hline
Configuration && C43-4 & C43-1 &  \multicolumn{2}{c}{7m}\\
Observation date && 21 Nov 2016 & 18 March 2017  & 12 Nov 2016 & 17 Nov 2016\\
Number of antennas && 41 & 43 & 12 & 11 \\
Number of pointings && 5 & 5 & 3 & 3\\
Time on source (mins) && 34 & 10 & 50 & 50\\
Flux calibrator && J1058+0133 & Ganymede & J1058+0133 & Ganymede\\
Bandpass calibrator && J1058+0133 & J1058+0133 & J1058+0133 & J1058+0133\\
Phase calibrator && J1103+3014 & J1150+2417& J1159+2914 & J1159+2914\\
 \hline
 \end{tabular}
\label{table:ObserPara}
\end{table*}

Data reduction was performed using the
Common Astronomy Software Applications (CASA) package 
\citep{mcmullin07},
version 4.7.0 for the C43-4 and ACA 7m data and version 4.7.2 for the C43-1 data.
The calibrated visibilities were concatenated using the task CONCAT in CASA 5.4 and the continuum subtraction was done using line-free channels with the task UVCONTSUB.
As a result, using the TCLEAN task with Briggs weighting (robust = 0.5), the achieved synthesized beam size of the CO (2--1) image is 0\farcs79 $\times$ 0\farcs64 (PA=26\fdg3) and the rms noise is $\sim$ 1\,mJy\,beam$^{-1}$ for the velocity resolution of 6.5\kms.

To assess whether there is missing flux in the CO (2--1) ALMA data, due to the lack of short-spacing information in the uv-plane, we compared the integrated line flux 
with the result reported  by \cite{planesas97} using the IRAM 30m telescope. \cite{planesas97} 
estimated an integrated flux of CO (2--1) of 1343\,Jy\kms\ within 27\arcsec\  and derived a molecular gas mass of $2.1\times10^9$\,M$_\odot$, assuming the CO-to-H$_2$ conversion factor $X_{10}=N({\rm H_2})/I_{\rm CO(1-0)}$ is $3.0\times10^{20}$\,cm$^{-2}$\,(K\,\kms)$^{-1}$, a line intensity ratio $\rm R_{21}$= $I_{\rm CO (2-1)}/I_{\rm CO (1-0)}$ of 0.8, and that the distance to the galaxy is 20.7\,Mpc.
 From our ALMA data, the integrated flux obtained within 27\arcsec\ is  $1235$\,Jy\kms. 
Given that the absolute flux density accuracy of ALMA is about 5 to 10 per cent at $\sim$230\,GHz (Section 10.2 in ALMA  Technical  Handbook), the results above may imply that most of the flux within 27\arcsec\ is recovered by ALMA.
However, the lack of a quoted uncertainty in the IRAM result prevents us from stating a more definite answer. For example, an error of 10 per cent in the IRAM flux estimate would make the two results
compatible within the uncertainties.

\section{Results}
\subsection{Continuum Emission}
\begin{figure*}
  \includegraphics[scale=0.43, clip=true, trim = 0 8 8 0]{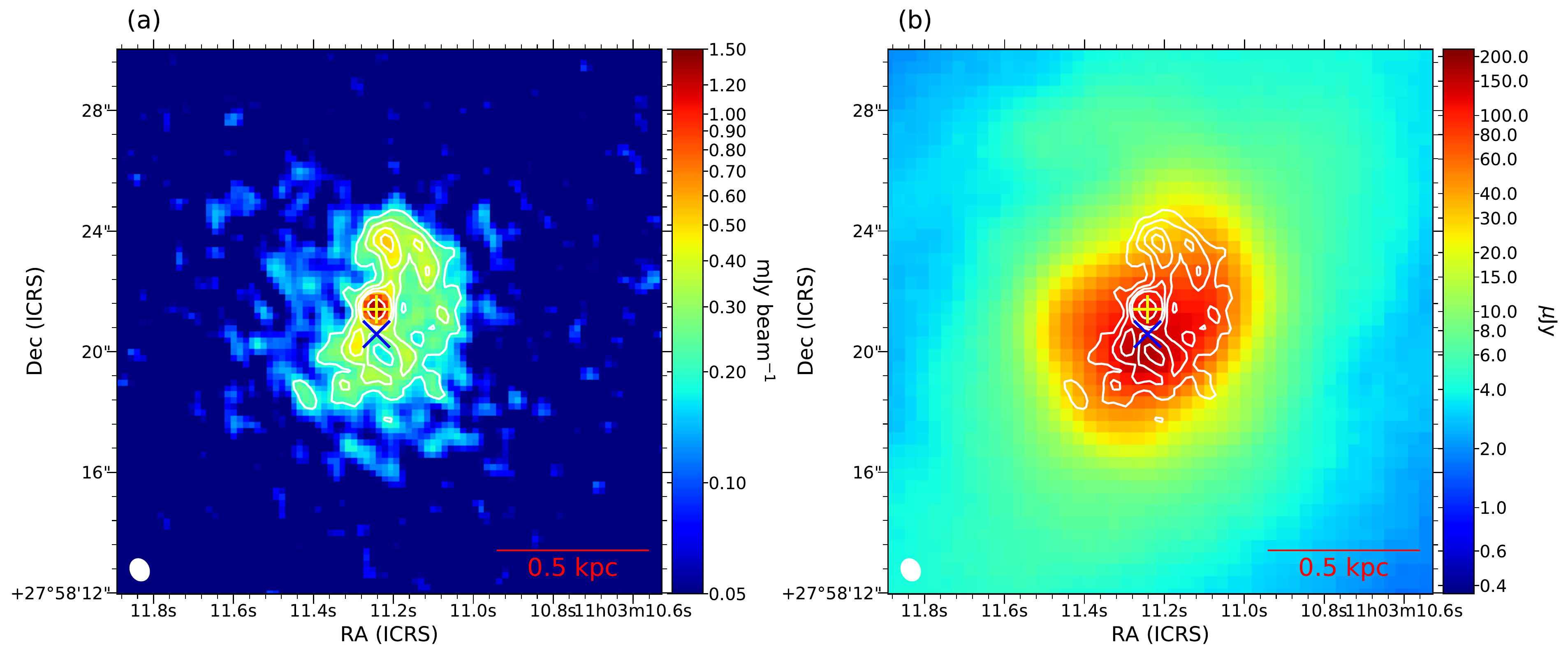}
 \caption{ (a) 230\,GHz continuum emission with contour levels of 5.0, 7.5, 10.0, 12.5, and 25.0$\sigma$, with $\sigma$=0.04 mJy beam$^{-1}$. 
 (b) 230\,GHz continuum overlaid on the SDSS $g$-band image of NGC\,3504. 
 Each image has a size of 18\arcsec $\times$ 18\arcsec\ with north up and east left.
 The synthesized beam (0\farcs75 $\times$ 0\farcs60, PA=25\fdg2) is shown on the lower-left corner of the map.
The yellow plus sign in each panel marks the center of the galaxy at (R.A., DEC) = ($11^{\rm{h}}03^{\rm{m}}11\fs24\pm0\fs002$, $+27\degr 58\arcmin 21\farcs 41\pm0\farcs03$) which is derived by two-dimensional Gaussian fitting. The blue cross sign indicates the galactic center location (R.A., DEC) = ($11^{\rm{h}}03^{\rm{m}}11\fs24$, $+27\degr 58\arcmin 20\farcs58$) given by SDSS.}
\label {fig:SDSS_Cont}
\end{figure*}

The 230\,GHz continuum image of NGC\,3504 was made using all four spectral windows after the lines were subtracted. The available bandwidth is about 6234\,MHz, yielding an rms noise level of 0.04\,mJy\,beam$^{-1}$, with a 0\farcs75 $\times$ 0\farcs60 (PA=25\fdg2) synthesized beam. 
Fig.~\ref{fig:SDSS_Cont} displays the 230\,GHz continuum image and the continuum contours superposed on the SDSS $g$-band image. 
It clearly shows that the continuum emission is asymmetric and is characterized by a bar-like structure with a compact nucleus, and mostly confined in a radius of 3\farcs5 ($\sim$ 0.35\,kpc). The peak flux is 1.45\,mJy\,beam$^{-1}$, which translates to a $\sim$ 36$\sigma$ significance.
The location of the galactic center (R.A., DEC) = ($11^{\rm{h}} 03^{\rm{m}} 11\fs24\pm0\fs002$, $+27\degr 58\arcmin 21\farcs41\pm0\farcs03$) is derived by two dimensional Gaussian fitting of the compact nucleus, as indicated by the plus sign in Fig.~\ref{fig:SDSS_Cont}.
That position can be compared with (R.A., DEC) = ($11^{\rm{h}} 03^{\rm{m}} 11\fs25$, $+27\degr 58\arcmin21\farcs0$), measured with 1.4\,GHz continuum emission by \citet{condon90a}.
The positions differ by $\sim~0\farcs4$,
smaller than the synthesized beam of 1\farcs5.
However, this position is about 0\farcs83 north of the galactic center as given by the SDSS, and is indicated by the blue cross sign in Fig.~\ref{fig:SDSS_Cont}. This difference is most likely due to the extinction by gas and dust at the galactic center in the SDSS data. 

\subsection{CO (2--1) Distribution}
\begin{figure*}
  \includegraphics[scale=0.35, clip=true, trim = 0 0 0 0]{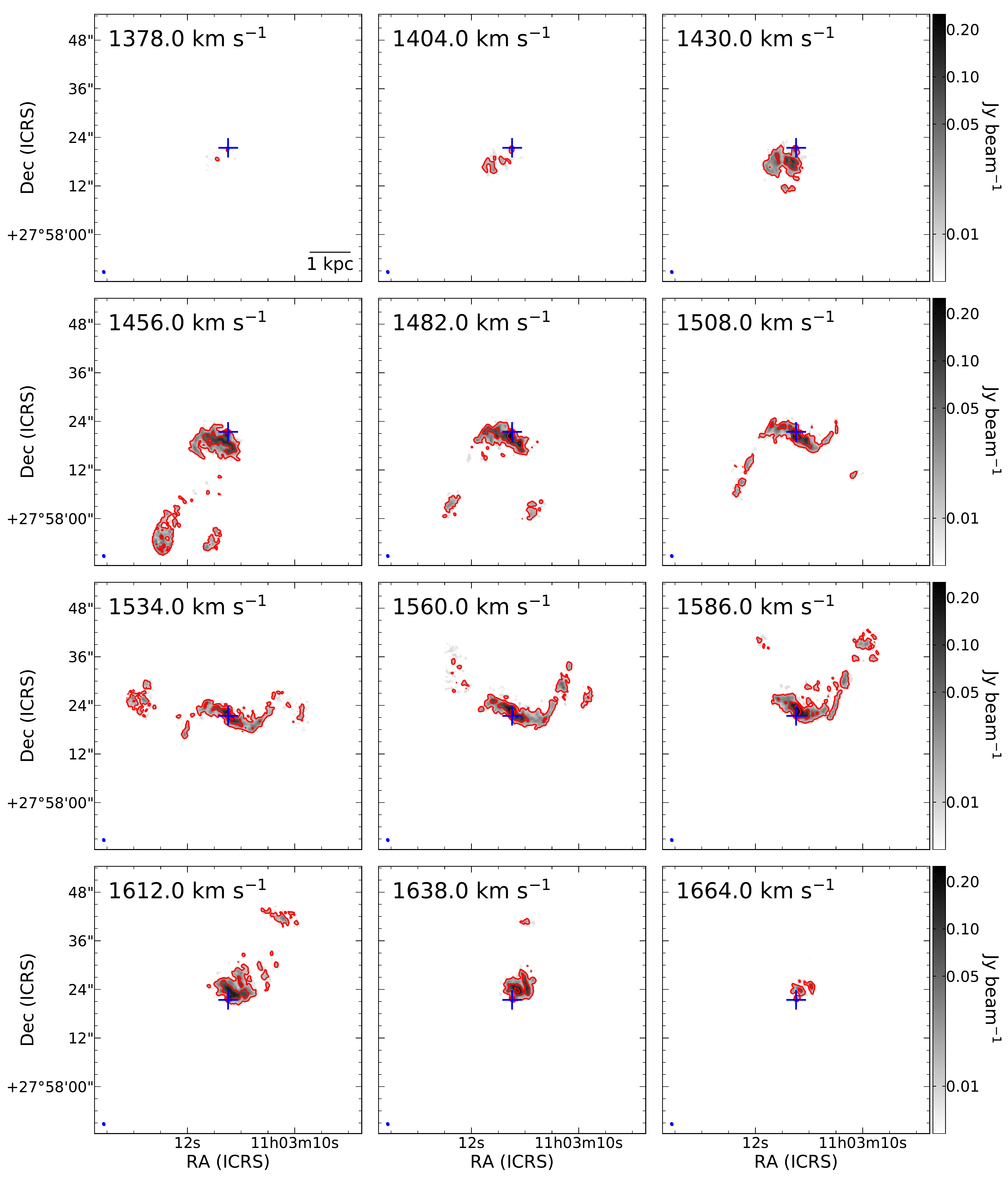}
 \caption{Channel maps of the CO (2--1) line of NGC\,3504. The velocities are shown at the upper-left corner of each panel. The synthesized beam (0\farcs79 $\times$ 0\farcs64, PA=26\fdg3) is shown on the lower-left corner of the map. Contour levels are 10, 50 and 200$\sigma$, which corresponds to 0.01, 0.05 and 0.20 Jy beam$^{-1}$. The plus sign in each panel marks the galactic center, as in Fig.~\ref{fig:SDSS_Cont}.}
\label {fig:channel_maps}
\end{figure*}

\begin{figure*}
 \includegraphics[scale=0.35, clip=true, trim = 0 0 0 0]{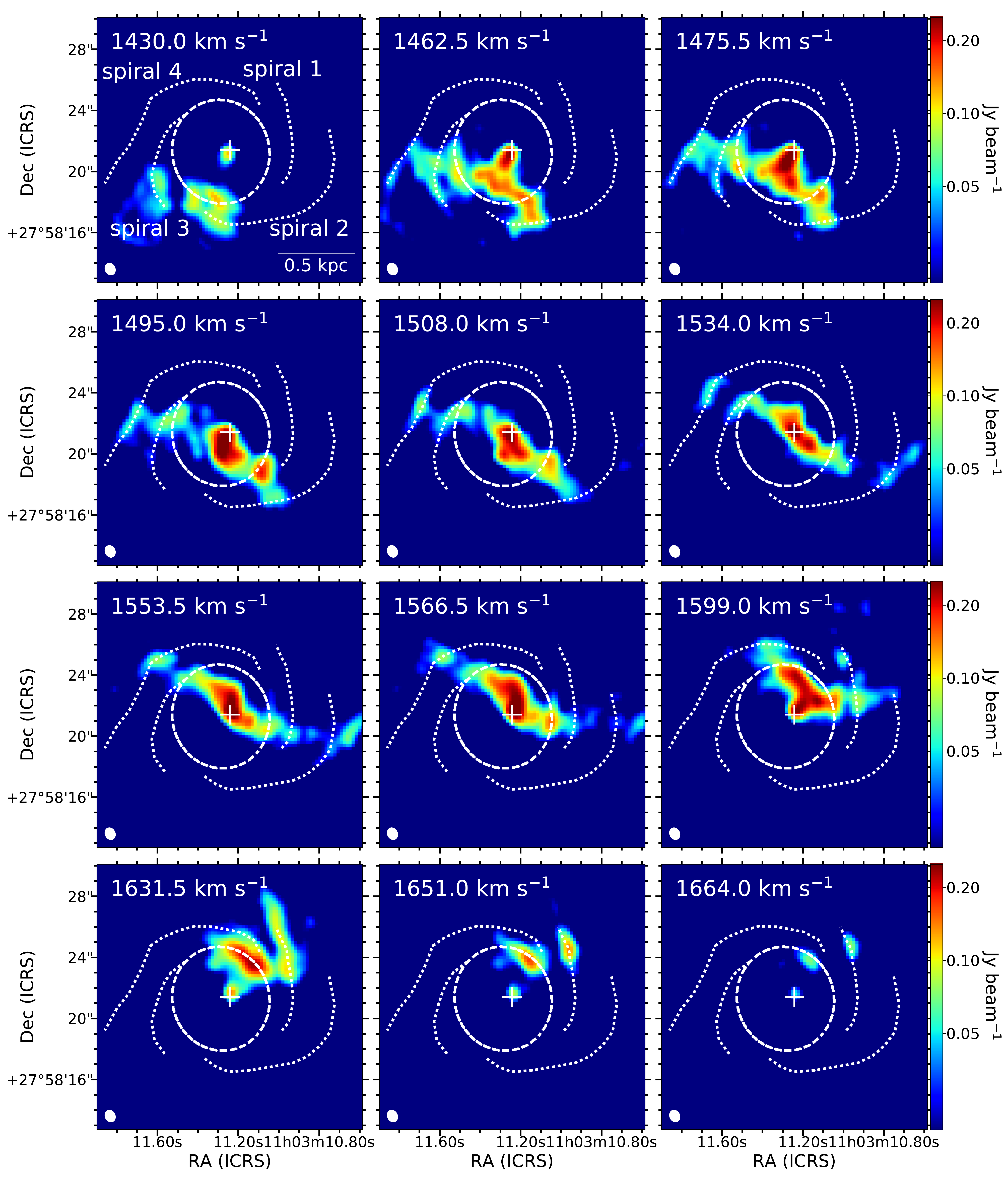}
 \caption{Channel maps of the CO (2--1) line of NGC\,3504 at the central $\sim$ 1.7\,kpc region. The velocities are shown at the upper-left corner of each panel. The dashed and dotted lines mark the identified nuclear ring and the four spiral structures, respectively. The synthesized beam is shown on the lower-left corner of the map. The plus sign in each panel marks the galactic center, as in Fig.~\ref{fig:channel_maps}.}
\label {fig:channel_maps_zoomin}
\end{figure*}

The CO (2--1) channel maps are shown in Fig.~\ref{fig:channel_maps}. 
There is no detection about 28\arcsec\ (2.8\,kpc) away to the southeast of the galactic center
because the mosaic observation did not fully cover this region. 
However, the observations did cover the dust lane in the outer bar region, which is within 28\arcsec\ southeast of the galactic center.
Fig.~\ref{fig:channel_maps_zoomin} shows the CO (2--1) channel maps at the central $\sim$1.7\,kpc region, 
from where we identified four spiral features and one nuclear ring structure by tracking the molecular gas, as shown by the white dotted and dashed lines, respectively. 

\begin{figure*}
  \includegraphics[scale=0.45, clip=true, trim = 0 10 0 5]{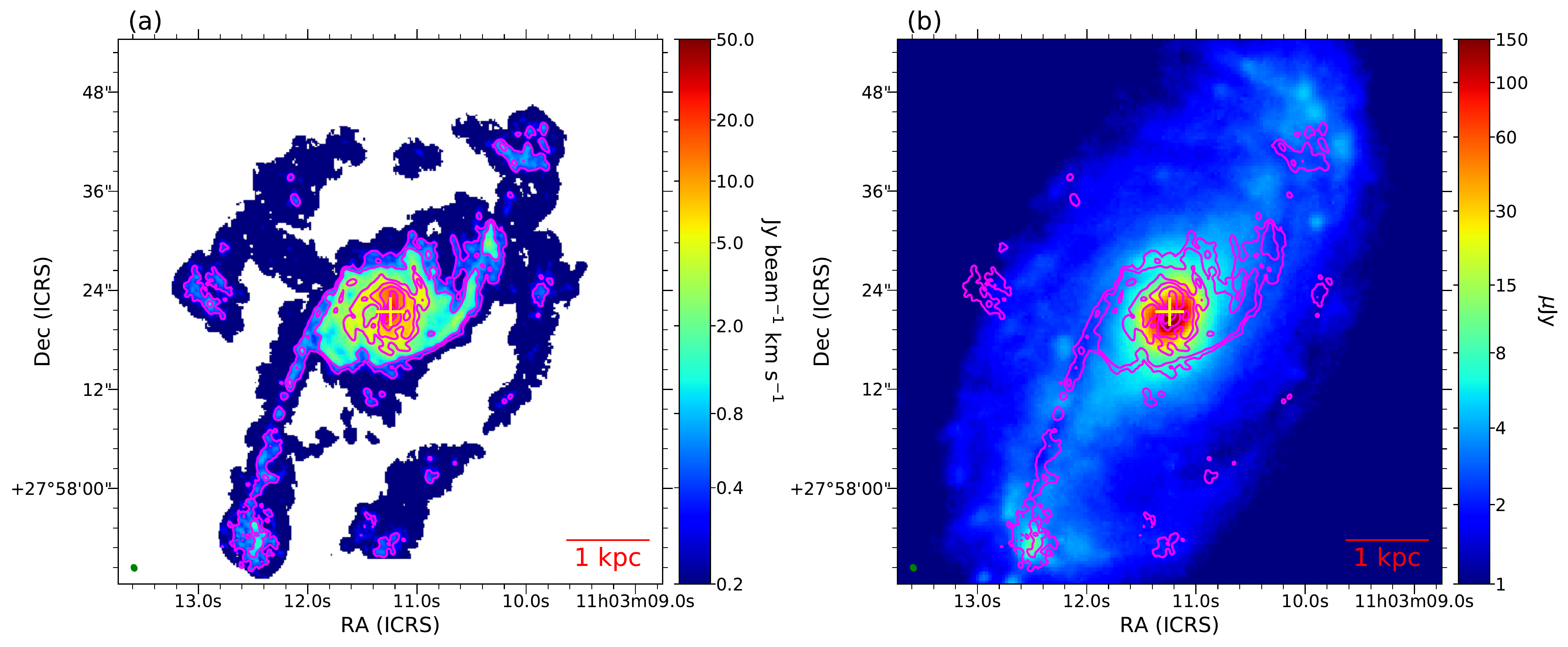}
 \caption{(a) CO (2--1) integrated intensity (moment 0) map. Contour levels are 0.35, 0.8, 2.0, 4.0, 10.0,15.0 and 30\,Jy\,beam$^{-1}$\kms. 
 (b) CO (2--1) contours from the moment 0 map overlayed on the SDSS $g$-band image of NGC\,3504.
 The synthesized beam is shown on the lower-left corner of the map and the plus sign marks the center of the galaxy, same as in Fig.~\ref{fig:SDSS_Cont}.}
\label {fig:mom0_map}
\end{figure*}

Fig.~\ref{fig:mom0_map} (a) and (b) present the CO (2--1) integrated intensity (moment 0) map and the CO (2--1) contours superimposed on the SDSS $g$-band image, respectively. It is clear that the CO (2--1) emission lies along the two dust lanes.  We also notice that the CO (2--1) emission on the two dust lanes are asymmetric in shape, and there are more massive clumps (integrated intensity larger than 2.0\,Jy\,beam$^{-1}$\kms) located on the northern dust lane.

\begin{figure*}
  \includegraphics[scale=0.69, clip=true, trim = 0 10 0 5]{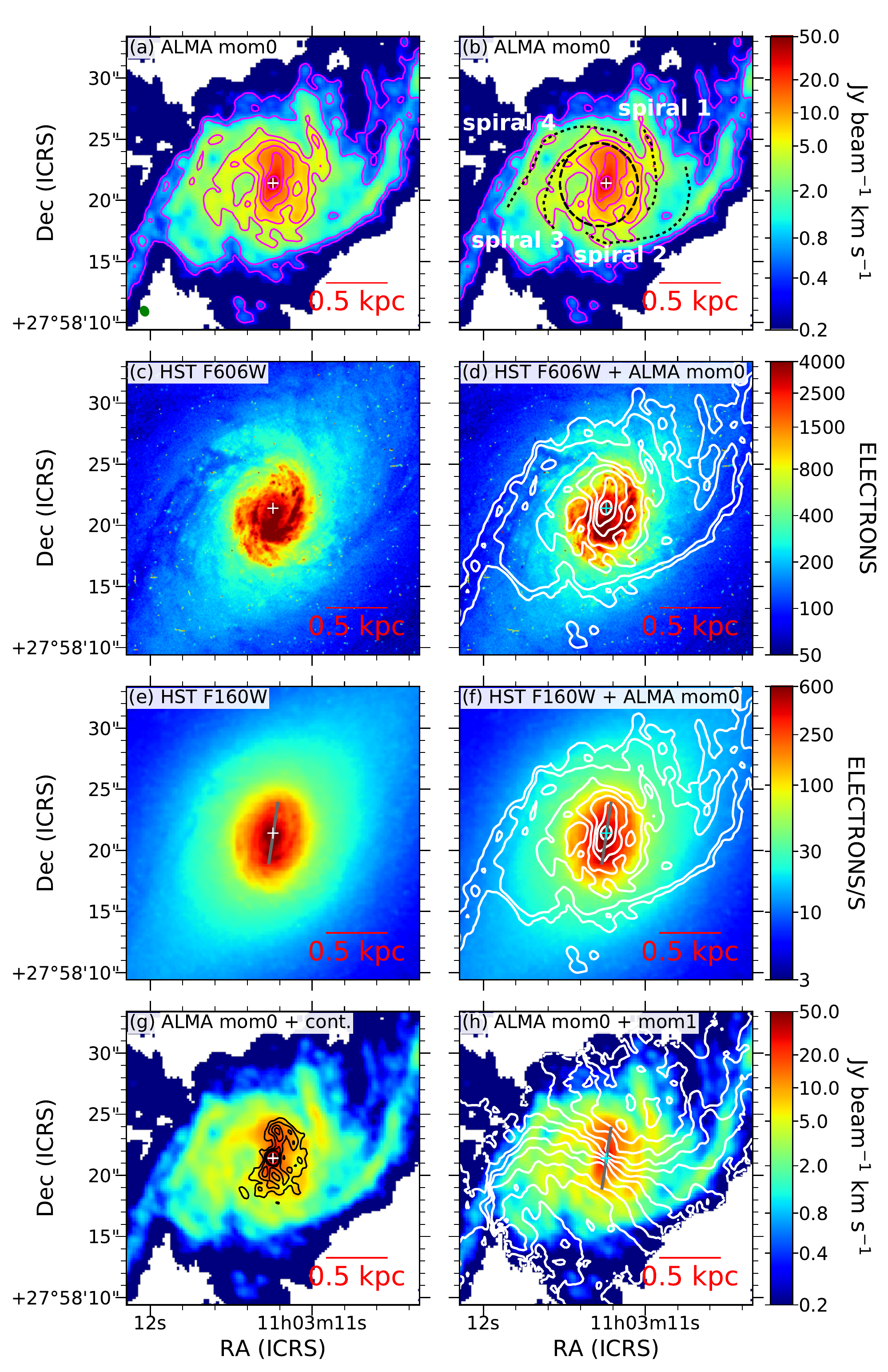}
 \caption{(a) and (b): CO (2--1) integrated intensity map of the central 2.4\,kpc region with the same contours as in Fig.~\ref{fig:mom0_map}.
 The black dashed and dotted lines in panel (b) are the identified nuclear ring and spiral structures, as in Fig.~\ref{fig:channel_maps_zoomin}.
  The synthesized beam is shown on the lower-left corner of the map. 
 (c) to (f): The HST/WFPC2 F606W and HST/WFC3 F160W images, together with the CO (2--1) contours. 
 (g) Superposition of the 230\,GHz continuum contours on the CO (2--1) integrated intensity map. The contour levels are the same as in Fig.~\ref{fig:SDSS_Cont}.
(h) CO (2--1) isovelocity contours superposed on the CO (2--1) integrated intensity map. 
The contour levels are from 1410 to 1670\kms\ in steps of 20\kms.
The grey line denotes the major axis of the inner bar, as in Fig.~\ref{fig:NGC3504}.
The plus signs mark the center of the galaxy, as in Fig.~\ref{fig:SDSS_Cont}.}
\label {fig:mom0_map_zoomin}
\end{figure*}

Fig.~\ref{fig:mom0_map_zoomin} (a) to (f) show the CO (2--1) integrated intensity (moment 0) map, together with the HST/WFPC2 F606W and HST/WFC3 F160W images, at the central 2.4\,kpc region.
An inner gas bar is present in NGC\,3504, as shown in Fig.~\ref{fig:mom0_map_zoomin} (a). 
The black dashed and dotted lines in Fig.~\ref{fig:mom0_map_zoomin} (b) indicate the identified nuclear ring and four spiral structures, as in Fig.~\ref{fig:channel_maps_zoomin}.
In Fig.~\ref{fig:mom0_map_zoomin} (d) we compare the CO (2--1) emission and HST/WFPC2 F606W optical image. 
It can be seen that the four identified spirals trace dust regions in the HST optical image, especially more clear for the spiral 4. 
In addition, the comparison of the CO (2--1) emission and HST/WFC3 F160W NIR image in Fig.~\ref{fig:mom0_map_zoomin} (f) shows that the position angle and the length of the inner gas bar are about the same as the inner stellar bar.
In Fig.~\ref{fig:mom0_map_zoomin} (g) we compare the CO (2--1) emission and the 230\,GHz continuum emission. 
It is apparent that the location of the continuum emission is consistent with the position of the strong CO (2--1) region. Because the 230\,GHz continuum mainly originates from the dust thermal emission, our result would imply that the dust is mixed with the gas near the galactic center, including at the inner gas bar.
Fig.~\ref{fig:mom0_map_zoomin} (h) presents the CO (2--1) isovelocity contours superposed on the CO (2--1) integrated intensity map. 
It can be seen that the velocity field slightly twists due to the inner gas bar. More details are described in Section 3.3.

\subsubsection{The Double-horned Profile}
\begin{center}
\begin{figure*}
  \includegraphics[scale=0.42, clip=true, trim = 0 10 0 5]{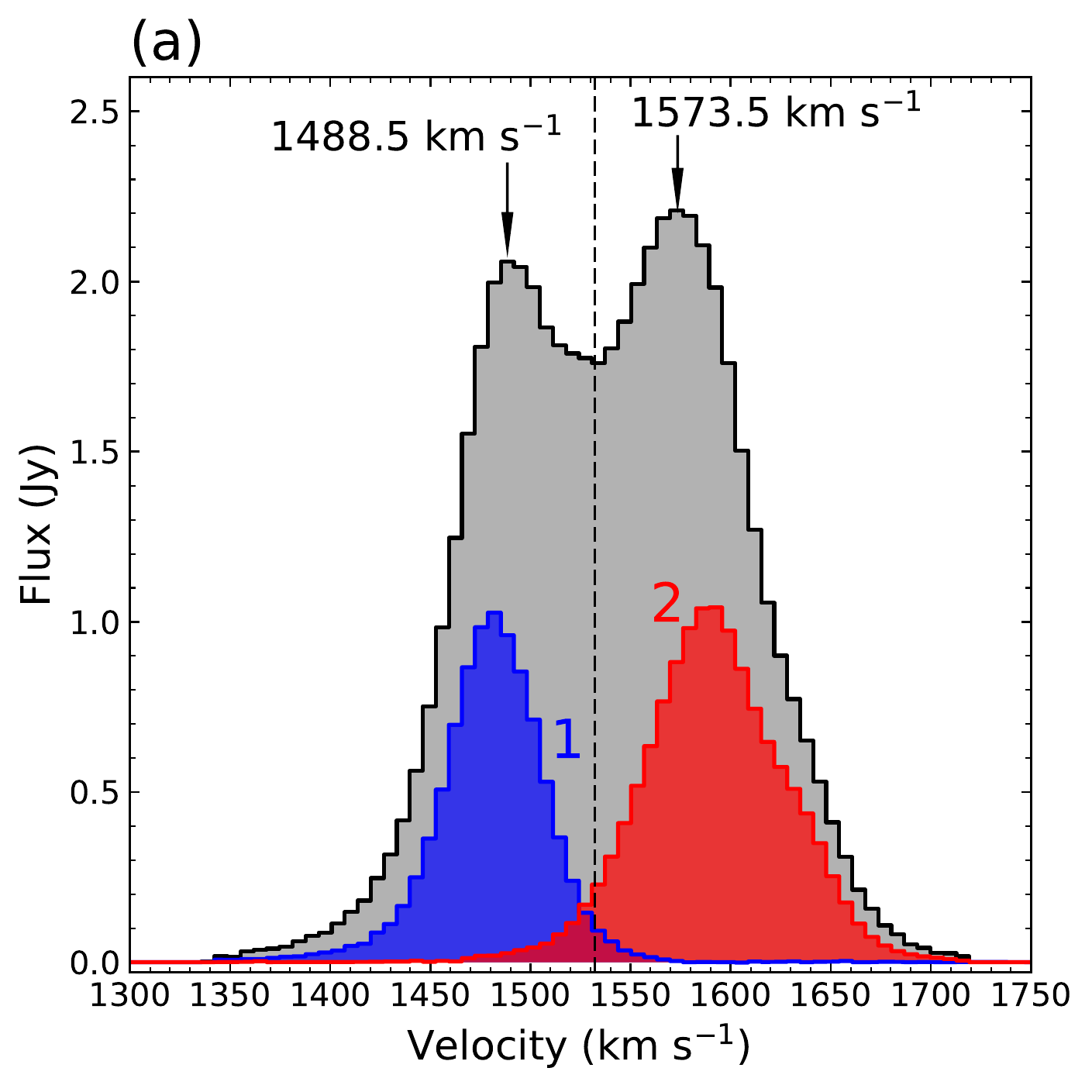}
  \includegraphics[scale=0.76, clip=true, trim = 0 5 0 5]{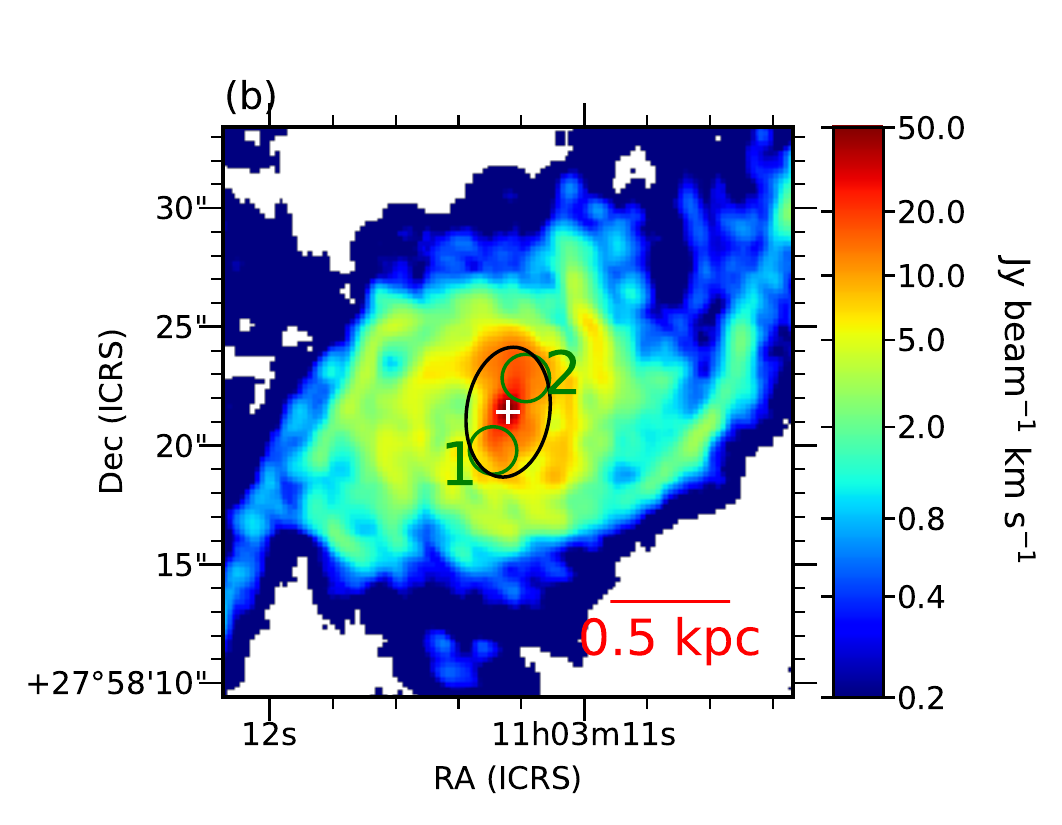}
 \caption{(a) CO (2--1) spectrum within the inner bar region (grey), region 1 (blue) and region 2 (red),
  as marked in panel (b).
 The vertical line indicates the systemic velocity of 1532.2\kms, as derived in Section 4.
(b) The black ellipse, with a major axis of 5\farcs5, a minor axis of 3\farcs5, and a position angle of 172\degr, indicates the inner bar region used for plotting the spectrum in panel (a). The green circles mark regions 1 and 2 with a radius of 1\arcsec.}
\label {fig:spectrum}
\end{figure*}
\end{center}

The total detected flux is 1480\,Jy\kms\ with $\sim$ 24 per cent within the inner bar region.
Fig.~\ref{fig:spectrum} (a) shows the spectrum over the inner bar region (in grey color).
The emission ranges from $\sim$ 1350\kms\ to $\sim$ 1715\kms.
There are two peaks at velocities of 1488.5 and 1573.0\kms, corresponding to peak fluxes of 2.0 and 2.2 Jy, respectively.
This double-horned profile was also found by \citet{planesas97} using the IRAM 30m telescope and noted that it could be due to an unresolved rotating ring-like structure or the concentration of molecular gas located at the end of a bar. 
However, \citet{planesas97} could not confirm the origin of this double-horned profile because the angular resolution of their data was insufficient to show a ring or a bar feature.
On the other hand, with the higher angular resolution of the ALMA data, we are able to see the double-horned profile from the inner bar region, as shown in Fig.~\ref{fig:spectrum}. 
In addition, within the inner bar region, about 50 per cent of the flux at velocities of 1488.5 and 1573.0\kms\ come from regions 1 and 2, respectively.

\subsubsection{Gas Mass}
The total mass of molecular gas is calculated using the equations (3) and (4) in \citet{solomon-vanden05} 
with the assumptions that the CO-to-H$_2$ conversion factor $X_{\rm 10}$ is $2 \times 10^{20}$\,cm$^{-2}$\,(K\kms)$^{-1}$ (corresponding to $\alpha_{\rm CO}=3.2$\,M$_{\odot}$\,pc$^{-2}$\,(K\kms)$^{-1}$), the line intensity ratio ${\rm R_{21}}=$  $I_{\rm CO (2-1)}/I_{\rm CO (1-0)}$ is 0.8, and the correction for the second most abundant element Helium is 1.36.
Therefore, the total molecular gas mass is 
\begin{equation}
{\rm M}_{\rm gas}[\rm M_\odot] =1.36 \, \alpha_{\rm CO} \, L^{\prime}_{\rm CO(2-1)}/R_{21},
\label{eq:gas_mass}
\end{equation}
where ${\rm L^{\prime}_{CO(2-1)}} = 3.25 \times 10^7 \, S_{\rm CO(2-1)} \, \Delta v \, \nu^{-2}_{\rm obs} \, D_L^2 \, (1+z)^{-3}$. 
The velocity-integrated flux $S_{\rm CO(2-1)} \, \Delta v$ is measured in Jy\kms, the observed frequency $\nu_{\rm obs}$ in GHz, the luminosity distance $D_L$ in Mpc, and $z$ is the redshift.
Given the total  estimation of flux of $1480$\,Jy\kms, the total molecular gas mass within our mapping area ($\sim50\arcsec \times 57\arcsec$) is $ 3.1 \times 10^9$\,M$_{\odot}$, which could be 5--10 per cent uncertain given the absolute flux accuracy alone, as mentioned in Section 2.

In addition, we calculated the concentration factor ($f_{\rm con}$) of NGC\,3504. The $f_{\rm con}$ is defined as the ratio of the surface density of molecular gas averaged over the central kiloparsec to the surface density of total molecular gas averaged over the whole optical disk at $R<R_{25}$ (the radius at a B-band surface brightness of 25 mag arcsec$^{-2}$) \citep{Sakamoto99}. Note that $R_{25}$ is 2\farcm5 in NGC\,3504 \citep{kenney93}. 
Given the averaged surface density at $R<500$\,pc and $R<2$\farcm5 of 1360.3\,M$_{\odot}$\,pc$^{-2}$ and 4.4\,M$_{\odot}$\,pc$^{-2}$, respectively, the concentration factor $f_{\rm con}$ of NGC\,3504 is 309.
This value is larger than 100.2$\pm$69.8, the average of 10 barred galaxies
reported in \cite{Sakamoto99}. This result agrees with \cite{Jogee2005J}, which shows that NGC\,3504 has a large central molecular gas concentration, comparable to other starburst galaxies and larger than typical barred galaxies.

\subsection{Kinematics}
\begin{figure*}
  \includegraphics[scale=0.43, clip=true, trim = 0 8 8 0]{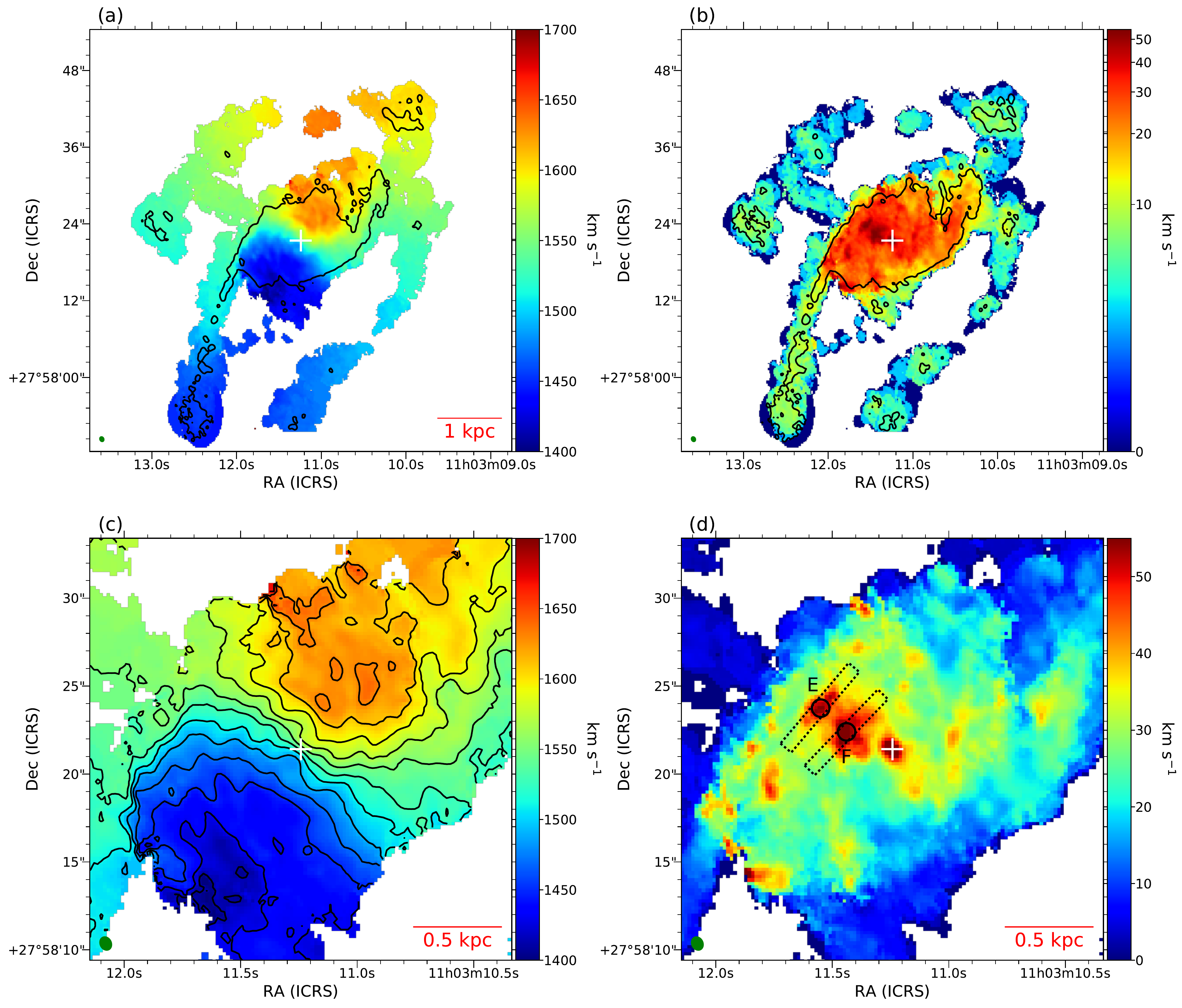}
 \caption{(a) CO (2--1) velocity field (moment 1) with the contour of the CO (2--1) integrated intensity at the level of 0.35 Jy\,beam$^{-1}$\kms.
 (b) CO (2--1) velocity dispersion image (moment 2) with the same contours as in panel (a).
 (c) Same as panel (a), but showing the central 2.4\,kpc region with contours from 1410 to 1670\kms\ in steps of 20\kms.
 (d) Same as panel (b), but showing the central 2.4\,kpc region. 
 The circles with a radius of 0\farcs5 indicate regions E and F, which have higher velocity dispersion.
The dotted rectangles represent the slices used for extracting the position–velocity diagrams shown in Fig.~\ref{fig:spectrum_checkVdisp} (a) and (b).
  The synthesized beam is shown on the lower-left corner of the map and the plus sign marks the center of the galaxy, as in Fig.~\ref{fig:SDSS_Cont}.}
\label {fig:mom1n2_maps}
\end{figure*}

The CO (2--1) velocity field and velocity dispersion are shown in  Fig.~\ref{fig:mom1n2_maps}. 
In Fig.~\ref{fig:mom1n2_maps} (a) and (b), the black contour shows the CO (2--1) integrated intensity at a level of 0.35\,Jy\,beam$^{-1}$\kms, which helps to indicate the location of the dust lanes as shown in Fig.~\ref{fig:mom0_map} (a).
As can be seen in Fig.~\ref{fig:mom1n2_maps} (a), the velocity field crossing the dust lanes shows a velocity gradient, which is predicted due to gas shocks along the leading edges of bars \citep{Sanders1980,Athanassoula92}.
The velocity dispersion along the southern and northern dust lanes is about 10 and 20\kms, respectively, as shown in Fig.~\ref{fig:mom1n2_maps} (b).

\begin{figure*}
  \includegraphics[scale=0.42, clip=true, trim = 0 8 8 0]{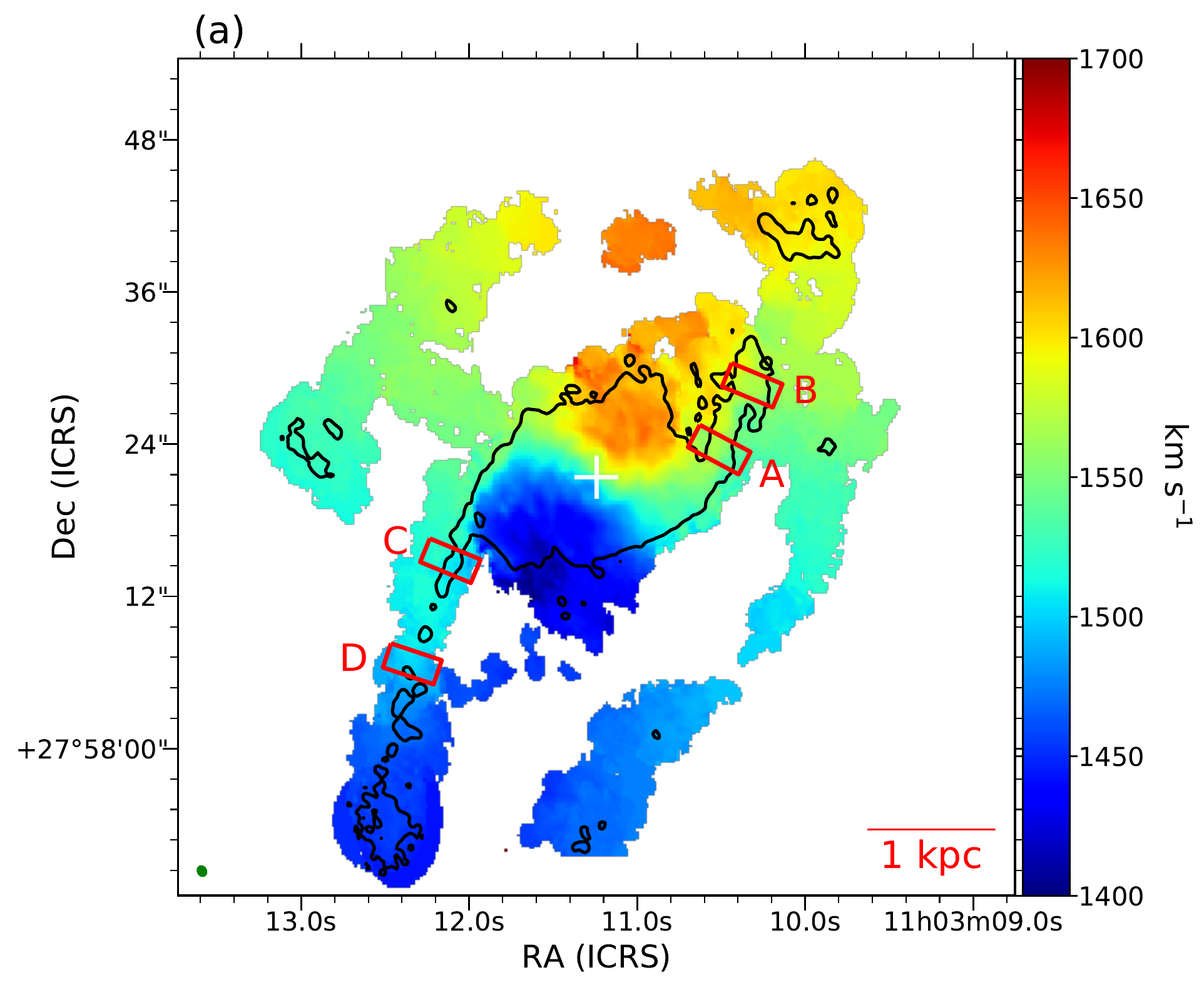}
    \includegraphics[scale=0.2, clip=true, trim = 0 8 8 0]{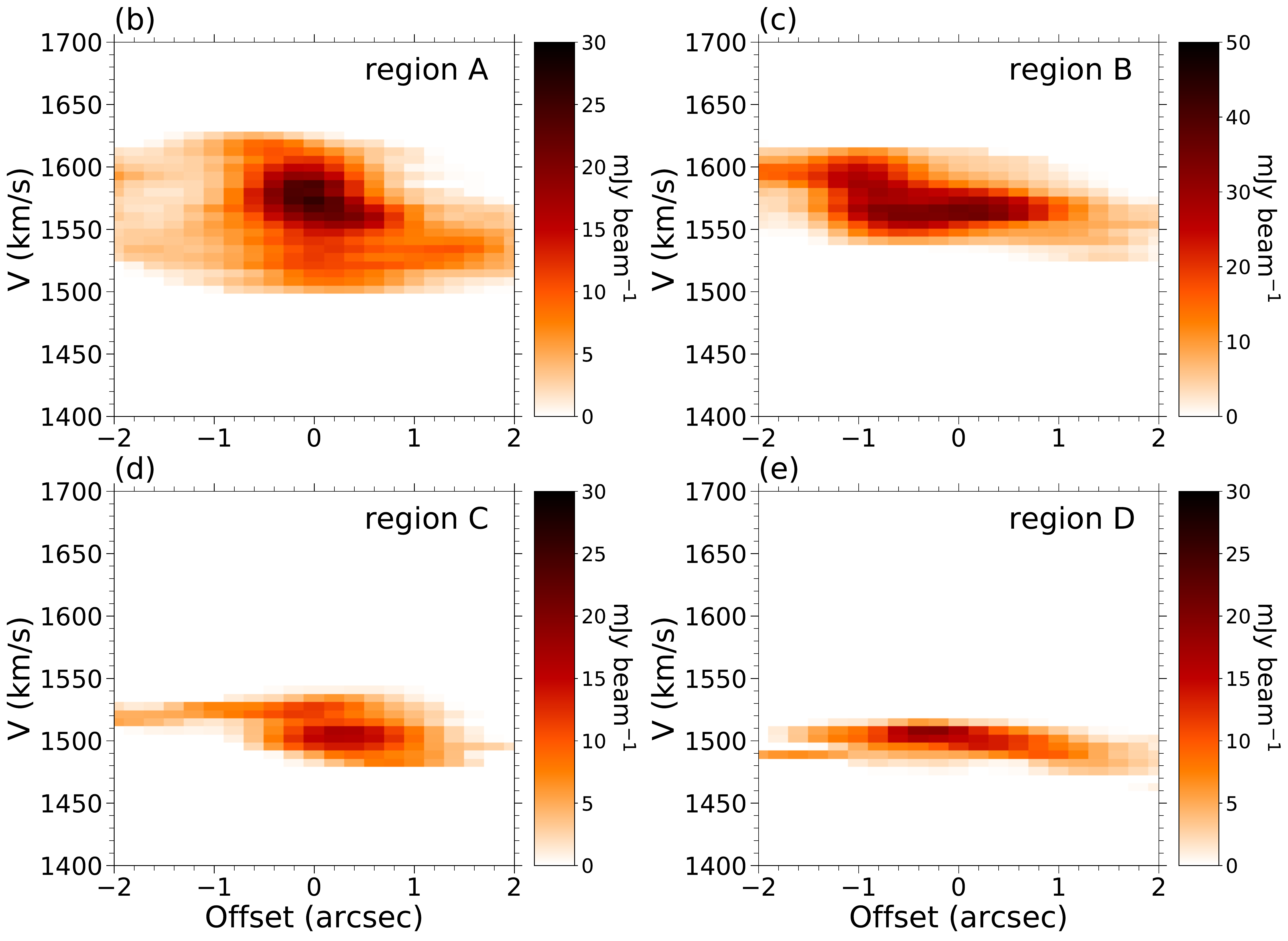}
 \caption{(a) Selected slices (A to D) perpendicular to the dust lanes with a width of 0.2 kpc overlaid on the CO (2--1) velocity field (moment 1). 
 The black contours of the CO (2--1) integrated intensity is at a level of 0.35\,Jy\,beam$^{-1}$, same as in Fig.~\ref{fig:mom1n2_maps} (a).
 (b) to (e): Position-velocity diagrams along slices (regions) A to D.
  }
\label {fig:PVD_DL}
\end{figure*}

To show more clearly the velocity gradient across the dust lanes, we made position-velocity (PV) diagrams along four slices (A to D) perpendicular to the dust lanes, as seen in Fig.~\ref{fig:PVD_DL}. 
The velocity gradient across the northern dust lane can be up to $\sim$50\kms\ over $\sim$ 1\arcsec, as shown in Fig.~\ref{fig:PVD_DL} (b) and (c).
On the other hand, panels (d) and (e) show that the velocity gradient can be up to about 30\kms\ over $\sim$ 1\arcsec\ across the southern dust lane.
After correcting for the inclination angle of $25\degr$, which is derived in Section 4, the corresponding velocity gradients are 1.2 and 0.7\kms\,pc$^{-1}$ across the northern and southern dust lane, respectively.

In addition to the molecular gas along the dust lanes, the CO (2--1) velocity field and velocity dispersion over the central 2.4\,kpc region are presented in Fig.~\ref{fig:mom1n2_maps} (c) and (d). 
The velocity contours in Fig. \ref{fig:mom1n2_maps} (c) are also overlaid on the CO (2--1) integrated intensity map in Fig.~\ref{fig:mom0_map_zoomin} (h). We find that the velocity field is twisted, likely due to the inner gas bar, and also slightly irregular due to the inner spiral structures and the dust lanes. 
No strong twist of the velocity field appears 3\arcsec\ North of the galactic center, which  implies that the possibility to have a massive second nucleus mentioned in \cite{Perez00} is low.
In addition, the steep velocity gradients of about 50\kms\ at the intersection of the dust lanes and the inner molecular region at $R\approx11\arcsec$ would imply the existence of shock fronts there.

\begin{figure*}
  \includegraphics[scale=0.365, clip=true, trim = 0 10 10 5]{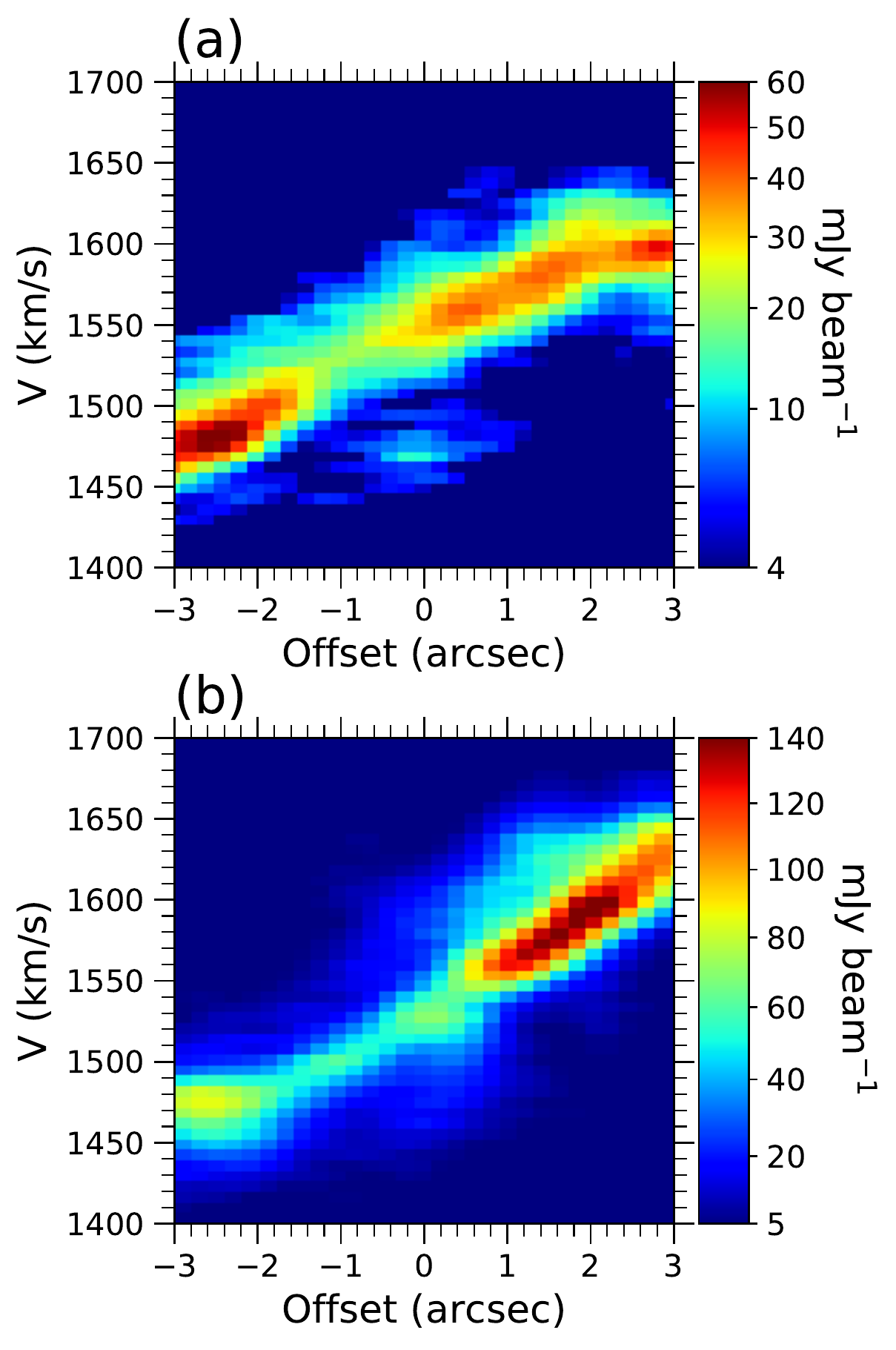}
    \includegraphics[scale=0.235, clip=true, trim = 0 0 0 0]{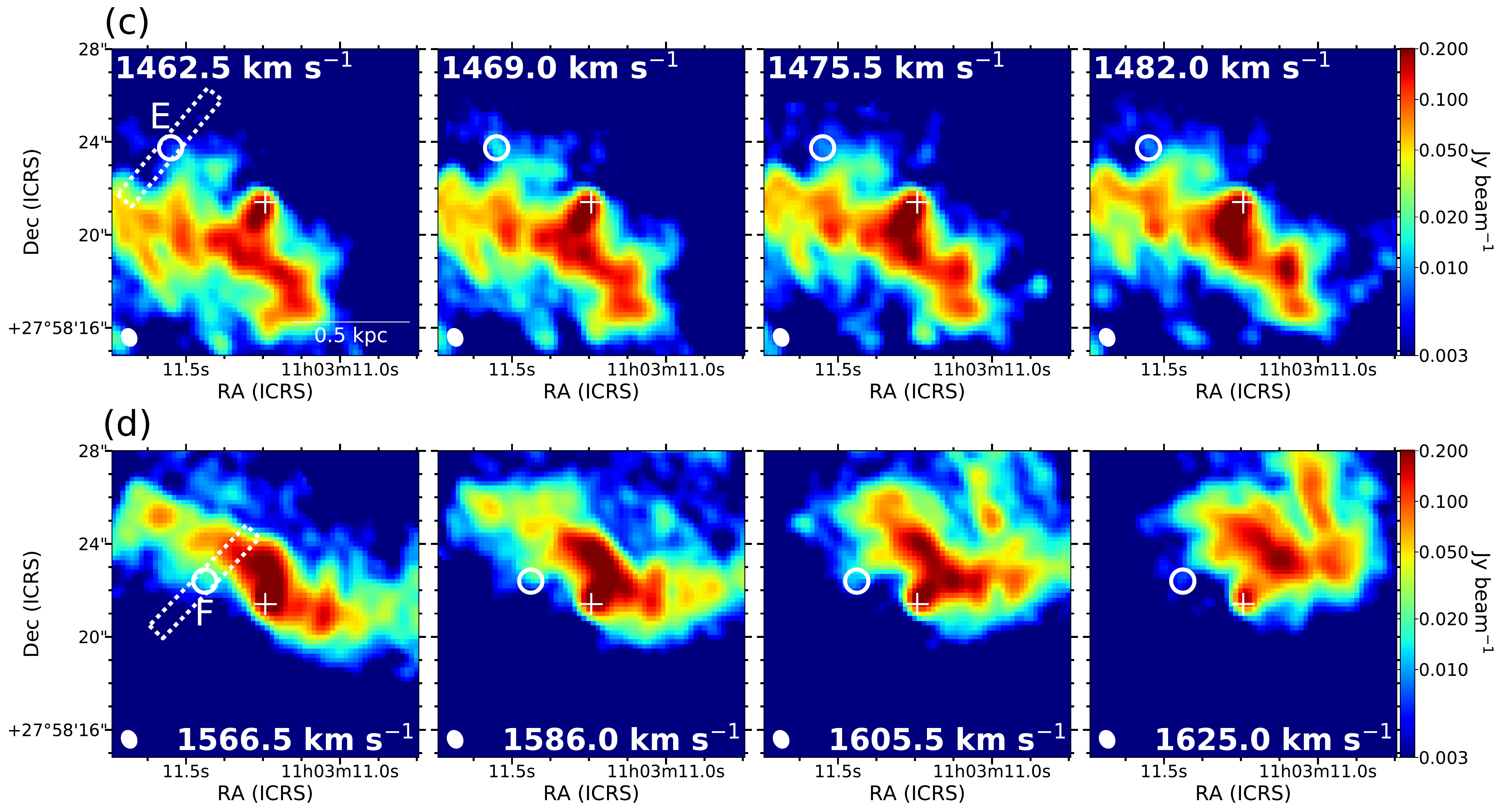}
 \caption{(a) and (b): 
 PV diagrams along two slices across regions E and F, as marked by the dotted rectangles in Fig.~\ref{fig:mom1n2_maps} (d).
(c) Channel maps of the CO (2--1) line of NGC\,3504 at the central $\sim$ 1.3\,kpc region.
 The velocity range is chosen to illustrate the origin of the CO (2--1) emission at $\sim 1470$\kms\ in panel (a).
 The white circle and the dotted rectangle mark region E.
 The synthesized beam is shown on the lower-left corner of the map. The plus sign marks the galactic center, as in Fig.~\ref{fig:channel_maps}.
 (d) Same as panel (c), but the velocity range is chosen to show the origin of the asymmetric CO (2--1) distribution around $\sim 1600$\kms\ in panel (b). The white circle and the dotted rectangle indicate region F.}
\label {fig:spectrum_checkVdisp}
\end{figure*}

The velocity dispersion increases to $\sim$ 30\kms\ in the central disk and to $\sim$ 55\kms\ at the galactic center as well as in the regions indicated by E and F in Fig.~\ref{fig:mom1n2_maps} (d). 
To probe the cause of the higher velocity dispersion in regions E and F, where the CO emission is relatively low ($\sim3\,$Jy\,beam$^{-1}$\kms), PV diagrams along them are shown in Fig.~\ref{fig:spectrum_checkVdisp} (a) and (b), respectively.
We can see that there is a discrete cloud with a velocity offset of $\sim70$\kms\ from most of the gas in region E.
This indicates that there is molecular gas at two distinct velocities at the same projected location, as opposed to one component with a broad line.
On the other hand, Fig.~\ref{fig:spectrum_checkVdisp} (b) shows an asymmetric broad component centered on the same velocity $\sim 1530$\kms\ in region F.
To reveal the origin of these two features in Fig.~\ref{fig:spectrum_checkVdisp} (a) and (b), 
Fig.~\ref{fig:spectrum_checkVdisp} (c) and (d) show the CO (2--1) channel maps at the central $\sim$ 1.3\,kpc region around $\sim 1470$\kms\ and $\sim 1600$\kms, respectively.
From Fig.~\ref{fig:spectrum_checkVdisp} (c), we can see that there is localized emission at $\sim 1470$\kms\, in panel (a) that is not connected to the main structure at higher velocities. Such emission is almost completely absent at other velocity frames and probably not related to the motion of the bar and spirals.
Also, Fig.~\ref{fig:spectrum_checkVdisp} (d) shows that the asymmetric distribution of the molecular gas around $\sim 1600$\kms\ in panel (b) is associated with a spur-like substructure at ($11^{\rm{h}} 03^{\rm{m}} 11\fs4$, $+27\degr58 \arcmin 23\farcs2$),
$\sim$ 2\farcs5 away to the northwest of the drawn circle. The feature is more clear in the channel maps at 1605.5 and 1625.0\kms.

\begin{figure*}
  \includegraphics[scale=0.34, trim = 0 0 0 0,clip=true]{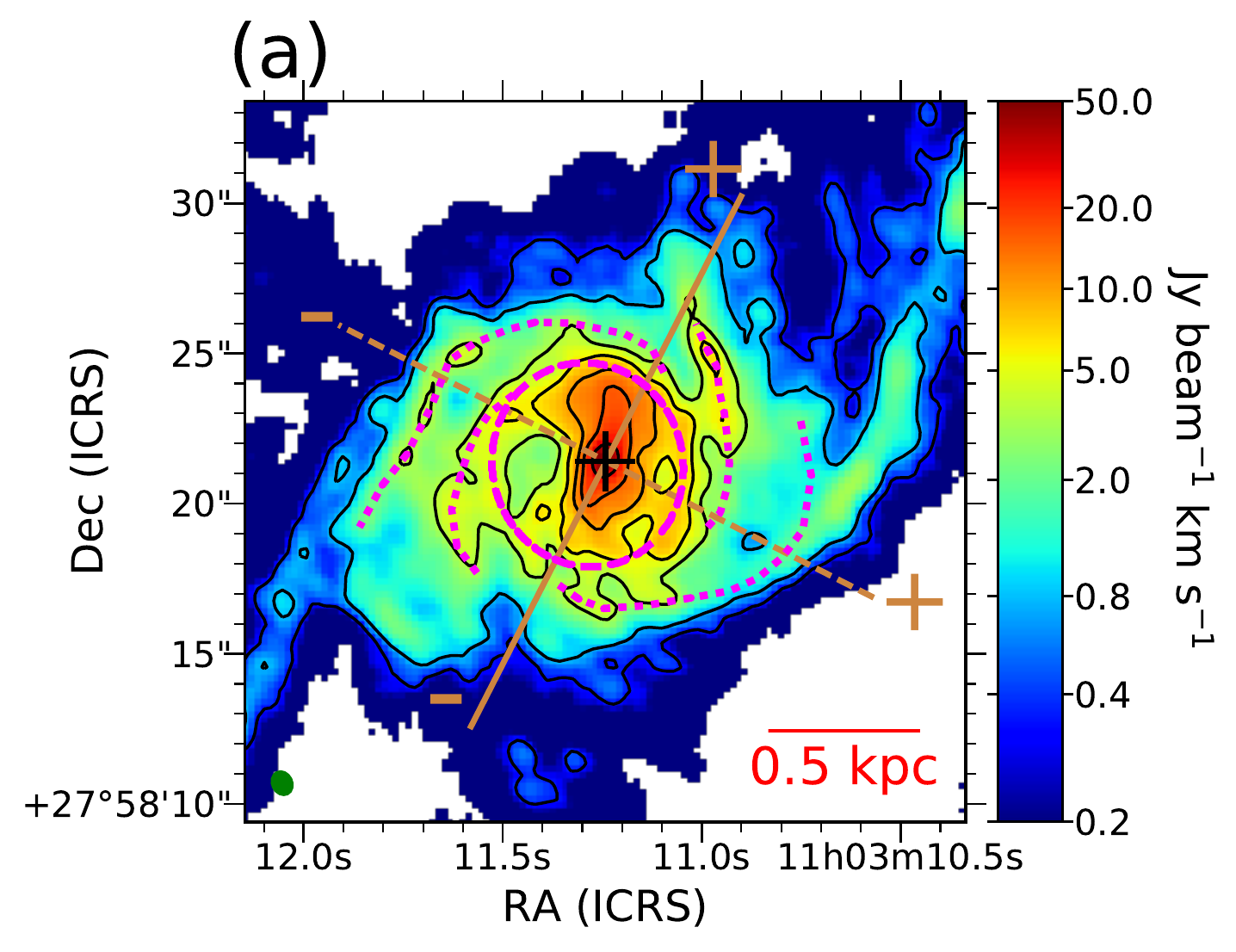}
  \includegraphics[scale=0.27, clip=true, trim = 0 0 0 0]{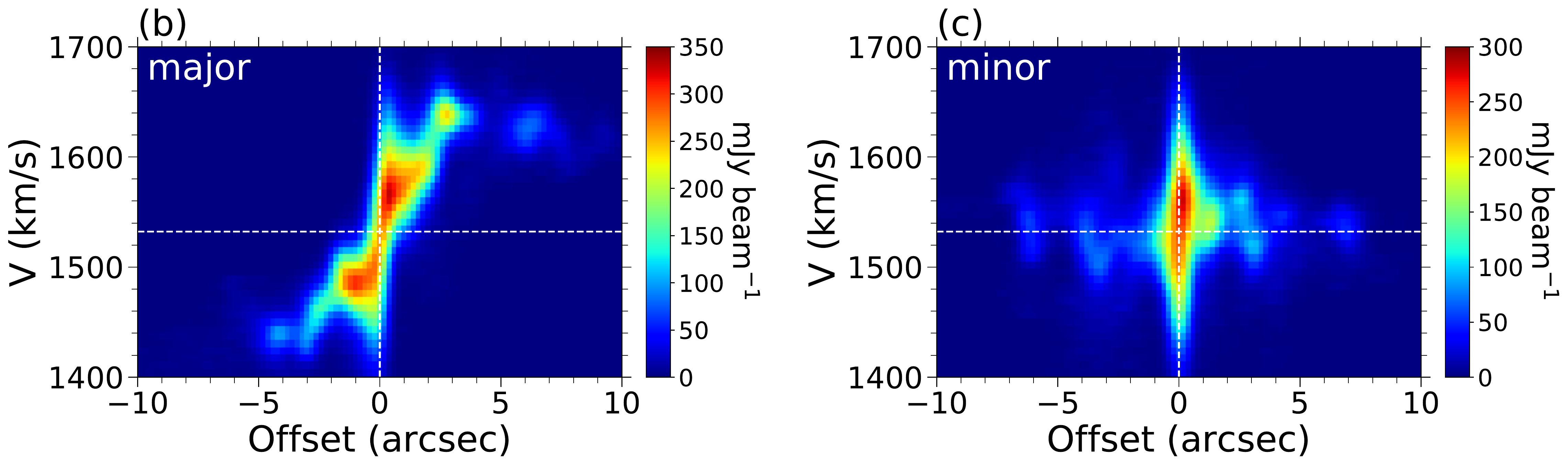}
 \caption{(a) CO (2--1) integrated intensity map of the central 2.4\,kpc region with the same contours as in Fig.~\ref{fig:mom0_map}. The brown solid and dashed lines indicate the slices along the major and minor axis of the galaxy, respectively. The plus and minus signs denote the direction of the offset in panels (b) and (c). The magenta dashed and dotted lines are the identified nuclear ring and spiral structures as in Fig.~\ref{fig:channel_maps_zoomin}.
 (b) and (c): PV diagrams along the major and minor axes with a width of the slice of 0\farcs8, same as the major axis of the synthesized beam. The horizontal dashed line is drawn at the system velocity of 1532.2\kms, as derived in Section 4.}
\label {fig:PVD_MajnMin}
\end{figure*}

Fig.~\ref{fig:PVD_MajnMin} shows the PV diagrams along the major and minor axes of the galaxy. The position angle of the galaxy is derived using {\it Kinemetry} \citep{kraj06} and will be described in Section 4. It is very close to the position angle derived from the optical isophotes in the outer disk \citep{grosbol85}. 
From Fig.~\ref{fig:PVD_MajnMin} (b), we can see a tilted bow-tie feature in the PV diagram along the major axis. This feature indicates the presence of two kinematically distinct gaseous components, which could be caused by a black hole or a non-axisymmetric potential, such as a bar \citep{Binney1991, Naray2009, Funes2002}. 
The velocity peaks located at offset positions of $+3\arcsec$ and $+6\arcsec$ are due to the presence of the nuclear ring and spiral 1 in the north, respectively.
The emission at offset position of $-4\arcsec$ is from spiral 2 located in the south.
Along the minor axis, as shown in Fig.~\ref{fig:PVD_MajnMin} (c), it can be seen that there exist non-circular motions. The non-circular motions within the central $\sim1\farcs5$ are an indication of the presence of an inner bar.
Outside the radius of $1\farcs5$, the wiggly patterns are mainly due to the presence of the nuclear ring and spirals.

\section{Kinematic Modelling}
To investigate the kinematics in NGC\,3504, we used {\it Kinemetry} \citep{kraj06} 
to fit our velocity (moment 1) field and quantify the contribution of circular rotation as well as  non-circular perturbations. 
The error of the velocity  field, $M_{\rm 1,err}$,  was calculated as
\begin{equation}
M_{\rm 1,err}=\sqrt{\sum_{i=1}^{n}\left(\frac{wv_i-\sum\limits_{j=1}^{n}\left(v_jI(v_j)\right)}{w^2}\right)^2\sigma_I^2+\sum\limits_{i=1}^{n}\frac{I^2(v_i)}{w^2}\sigma_v^2},
\end{equation}
where n is the number of channels, $I(v_i)$ is the intensity of a given pixel at a channel with velocity $v_i$, $w$ is the sum of the intensity $\sum\limits_{i=1}^{n}I(v_i)$, $\sigma_I$ is the standard deviation of the intensity at line-free channels, and $\sigma_v$ is the velocity uncertainty, which is assumed to be one half of our velocity resolution.

Under the assumptions that the velocity field can be decomposed  in a number of ellipses and the velocity profiles along the ellipses can be expressed as a Fourier series (see Equation (6) in \citet{kraj06}),
{\it Kinemetry} derives the system velocity, the position angle and the inclination of each ellipse by minimizing the coefficients of the Fourier series which are used to describe the non-circular velocity, as shown in Equations (7) and (8) in \citet{kraj06}.
In our case, we only considered six odd terms during the fitting, i.e. $A_n \sin(\psi)$ and $B_n \cos(\psi)$ with n=1, 3, 5, where $\psi$ is the eccentric anomaly. The even terms are usually very small and can be negligible in the velocity field \citep{kraj06,kraj08}.

In the first {\it Kinemetry} run we only fixed the galactic center to (R.A., DEC) = ($11^{\rm{h}} 03^{\rm{m}} 11\fs24$, $+27\degr58 \arcmin 21\farcs41$), which  was derived in Section 3.1, and obtained the system velocity of each ellipse. 
Then, in the second {\it Kinemetry} run, we fixed the velocity of the galaxy to the median value of the system velocities calculated from the first run and obtained the position angle (PA) and the inclination of each ellipse.
Finally, we fixed both PA and inclination to their median values and derive the coefficients of the Fourier series of each ellipse. 

Using the procedure described above, our best fit results yield a systemic velocity of the galaxy of
$1532.2 \pm 0.2$\kms, an inclination angle of $25\degr \pm 1\degr$, and a PA of $153\degr \pm 2\degr$, where the errors are the statistical errors corresponding to 1$\sigma$ confidence levels estimated by {\it Kinemetry} using Monte Carlo simulations. 
Our fit results for the galaxy velocity, inclination and PA are consistent with the values found by 
\citet{grosbol85} and \citet{kenney93},  $1535\pm2$\kms, 22\degr, PA=149\degr\ derived from the optical isophotes and PA=$147\degr \pm 2\degr$ derived from the nuclear CO (1--0) velocity field, respectively.

\begin{figure}
  \includegraphics[scale=0.6, clip=true, trim = 0 7 0 0]{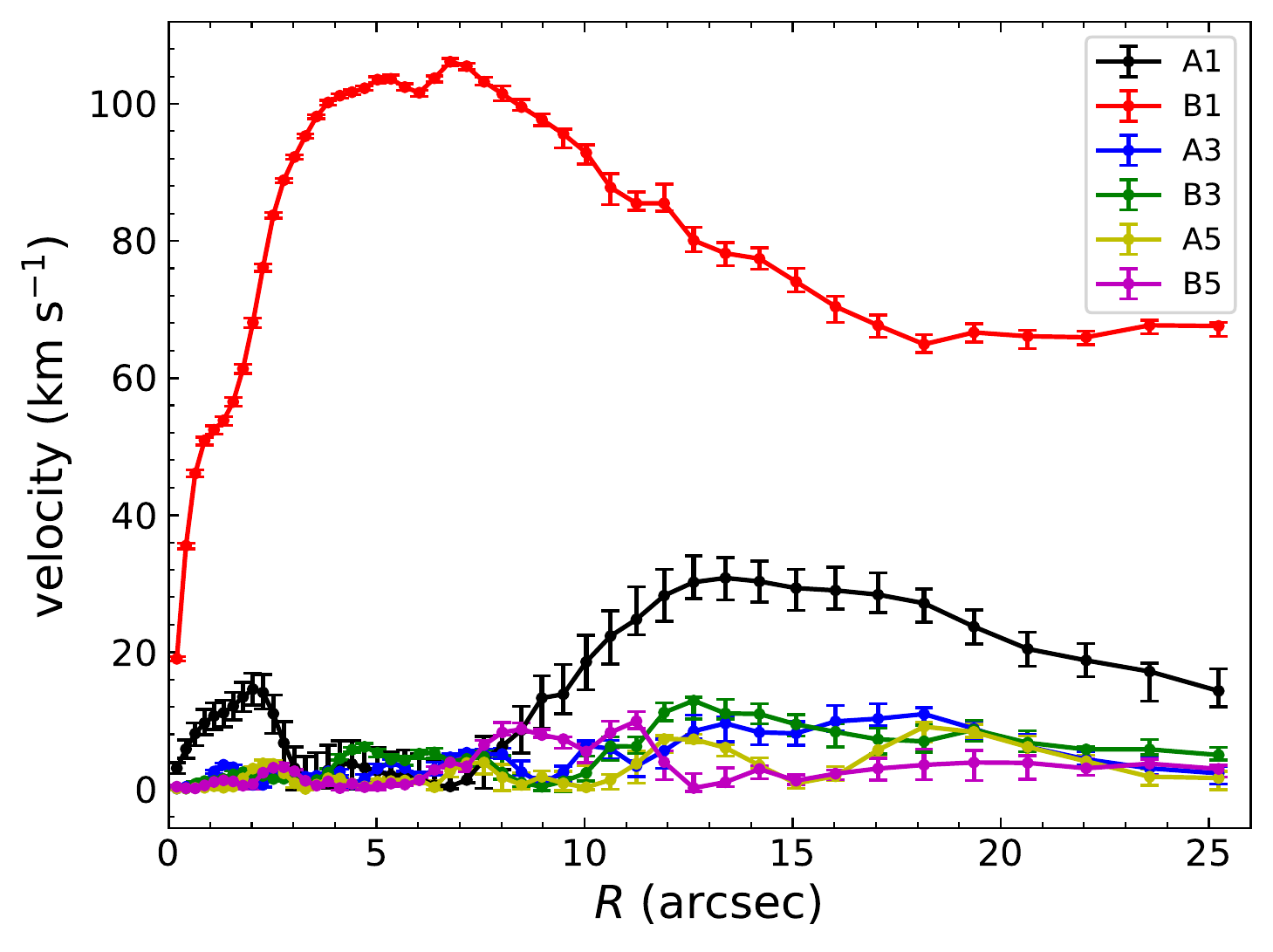}
 \caption{Coefficients of the Fourier components obtained from the $Kinemetry$ analysis as a function of the semi-major axis length of the ellipses.}
\label {fig:model_coefficient}
\end{figure}

Fig.~\ref{fig:model_coefficient} shows the Fourier decomposition coefficients of our fitting model as a function of the semi-major axis length of the ellipses. 
The errors indicate the variance caused by the uncertainty of the inclination and PA. 
The $B_1$ term, which represents pure circular motion, dominates in  most regions at $R<25\arcsec$.
The $A_1$ term, which indicates the existence of radial motion, is the second dominant mode in the innermost $\sim 3\arcsec$ and at $R>10\arcsec$.
Furthermore, the $A_3$ and $B_3$ terms, which are more sensitive to the center, inclination and position angle of the ellipses, and might also indicate complex velocity structures, are relatively small in our fit results.
The $A_5$ and $B_5$ terms are also small, and indicate the absence of strong multiple kinematic components, such as a counter-rotating disk, in the inner regions of NGC\,3504.

\begin{figure}
  \includegraphics[scale=0.5, clip=true, trim = 0 8 0 0]{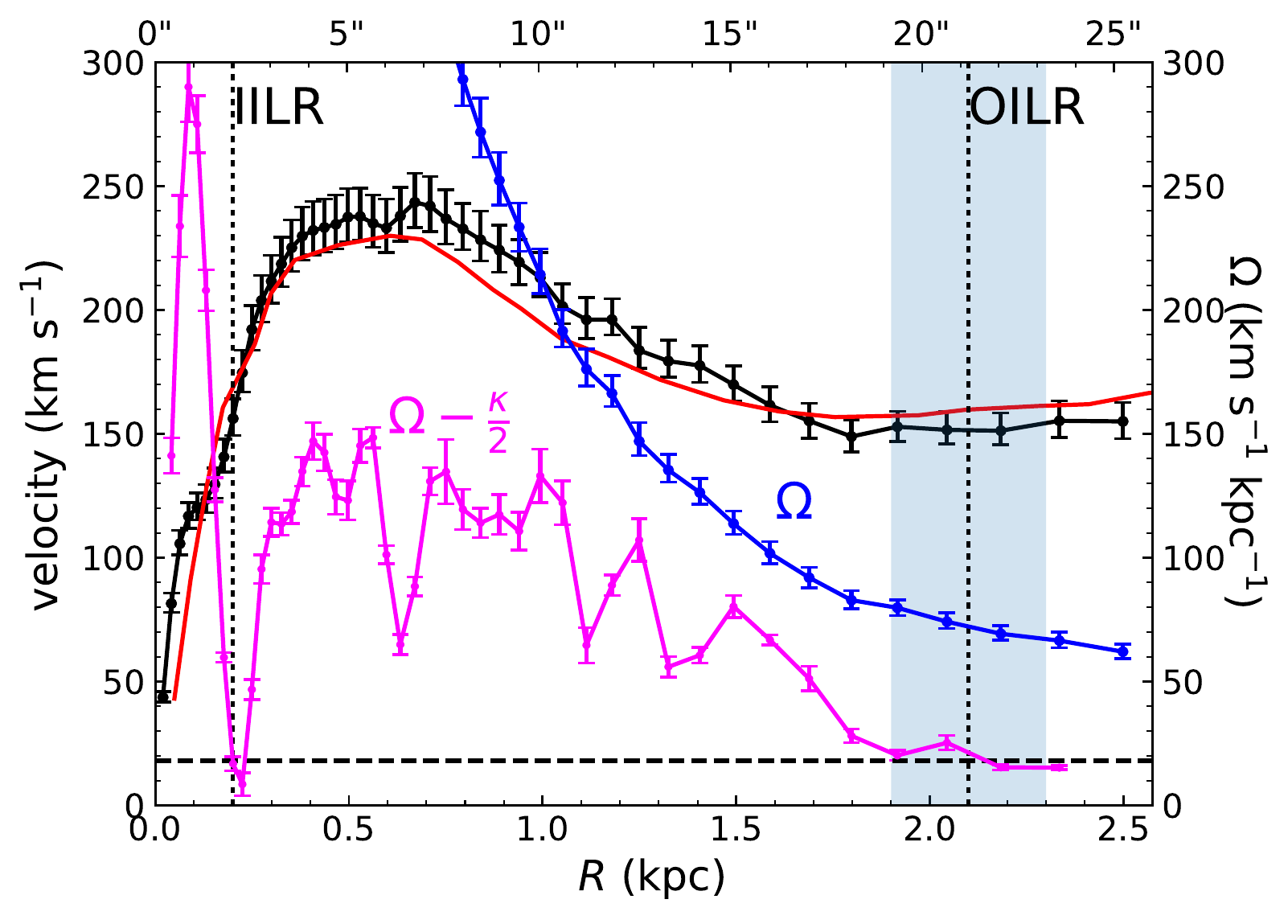}
 \caption{CO (2--1) rotation curve (black solid line) and angular velocity $\Omega$ (blue line) derived from our $Kinemetry$ analysis and corrected for the inclination of $i=25\degr$. The red line shows the rotation curve obtained by \citet{kuno00} from CO (1--0) and H$\alpha$ data and then corrected for the adopted inclination of $i=25\degr$. The black dashed line represents the derived pattern speed of the bar $\Omega_{\rm Bar}=18$\kmskpc\ in Section 6. With this pattern speed, an inner inner Lindblad resonance (IILR) is located at $R\sim 0.2$\,kpc and an outer inner Lindblad resonance (OILR) is predicted to exist between 1.9 and 2.3\,kpc, and are indicated by vertical black dotted lines.}
\label {fig:model_rot_curve}
\end{figure}

The rotation curve and the angular velocity $\Omega$ are presented by the black and blue lines in Fig.~\ref{fig:model_rot_curve}.
The rotational velocity was calculated from the $B_1$ term and corrected for inclination,  i.e. $v_\phi(R)=B_1(R)/\sin i$. The angular velocity was calculated as $\Omega(R) = v_\phi(R)/R$.
As can be seen from Fig.~\ref{fig:model_rot_curve}, the rotational velocity increases to $\sim 240$\kms\ at $R\sim 0.5$\,kpc and becomes nearly flat at $R>1.7$\,kpc. 
This is to be compared with the rotation curve derived by \citet{kuno00} using CO (1--0) (at $R<10\arcsec$) and H$_\alpha$ (at $R>15\arcsec$) data, and  corrected here for
 our adopted inclination angle of $i=25\degr$,
as shown by the red line in Fig.~\ref{fig:model_rot_curve}. We found that the two rotation curves are generally consistent with each other within $\sim$10\kms. 
Between $R\sim10\arcsec$ and 14\arcsec, our rotation velocities are slightly higher than those derived by \citet{kuno00}. This discrepancy could be due to the lack of data points at $R=10\arcsec - 15\arcsec$ in \citet{kuno00} and the use of different tracers, H$_\alpha$ and CO (1--0).

\begin{center}
\begin{figure*}
  \includegraphics[scale=0.43, clip=true, trim = 0 8 8 5]{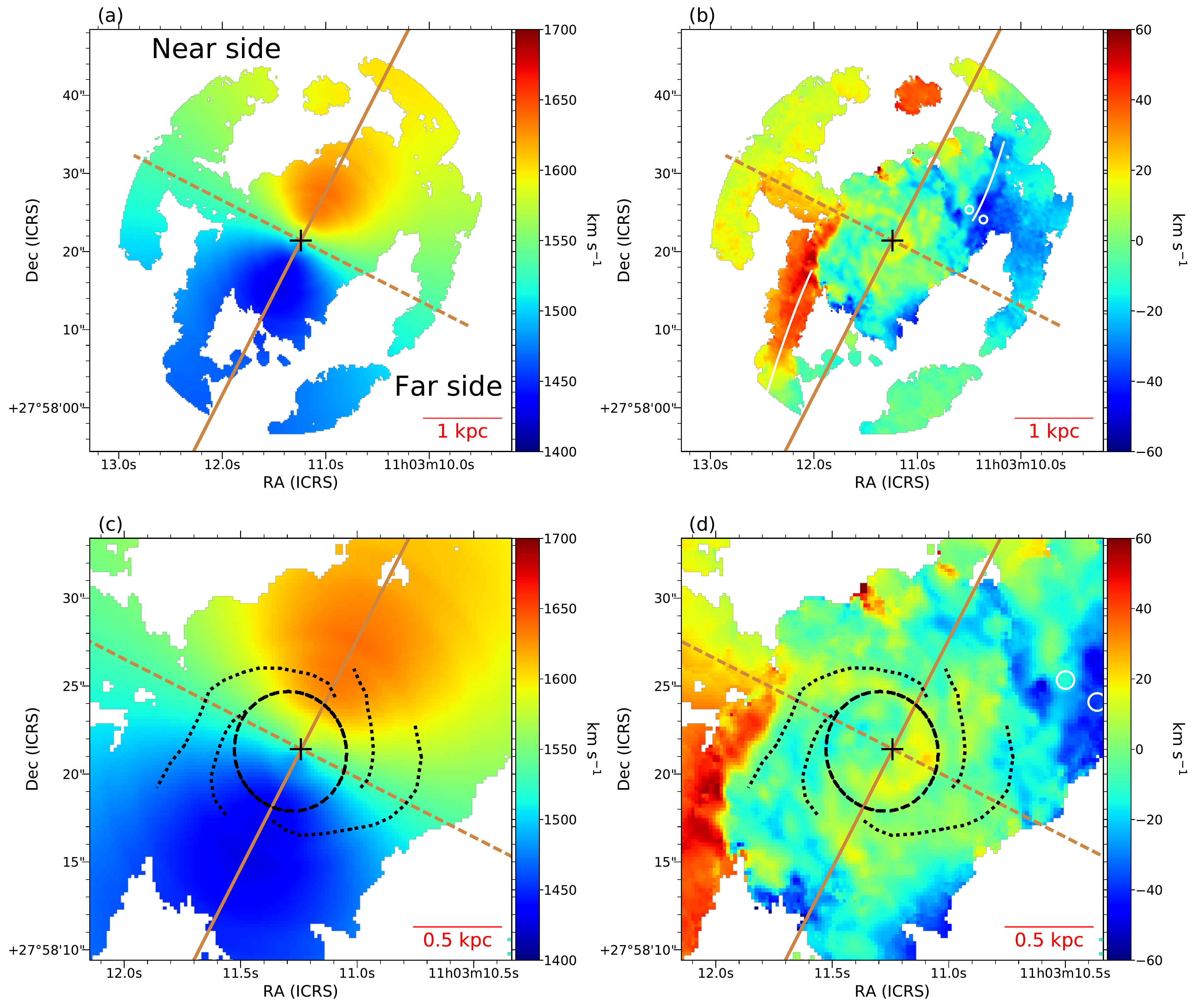}
 \caption{(a) Modelled velocity field generated by the term $A_0+B_1\cos(\psi)$ derived from $Kinemetry$ and then corrected for the inclination and PA. (b) Residual velocity map after subtracting the modelled velocity field from the CO (2--1) velocity field. The white lines indicates the two dust lanes identified on the SDSS $g-$band image (described in Section 5 and in Fig.~\ref{fig:Vst_R} (a)) and the two white circles mark the locations where the difference of the residual velocities is $\approx 30$\kms\ across the northern dust lane. The plus sign marks the center of the galaxy, as in Fig.~\ref{fig:SDSS_Cont}. The brown solid and dashed lines represent the galactic major and minor axes, respectively. (c) Same as panel (a), but showing the central 2.4\,kpc region. The black dashed and dotted lines are the identified nuclear ring and spiral structures as in Fig.~\ref{fig:channel_maps_zoomin}. (d) Same as panel (b), but showing the central 2.4\,kpc region.}
\label {fig:model_res_maps}
\end{figure*}
\end{center}

Using the derived parameter $B_1$ as well as the systemic velocity, the inclination and PA, we can generate the modelled velocity field from the derived rotation curve, as shown in Fig.~\ref{fig:model_res_maps} (a) and (c). The difference between the ALMA CO (2--1) velocity field and the  modelled velocity field is presented in Fig.~\ref{fig:model_res_maps} (b) and (d). Assuming that the dust lanes are located at the leading side of the bar \citep{Athanassoula92}, we can identify the far and near sides of the galaxy from Fig.~\ref{fig:model_res_maps} (a). From Fig.~\ref{fig:model_res_maps} (b), we also see that the residual velocities can reach $|v|\approx 50$\kms\ along the two dust lanes and are different across them.
The difference of the residual velocity across the dust lanes on the leading and trailing side of the bar
can be up to 
$\approx 30$\kms\ 
when we compare the residual velocity at the locations indicated by two circles in Fig.~\ref{fig:model_res_maps} (b).
Furthermore, the absolute value of the residual velocity is larger across the dust lanes on the leading side of the bar. This implies that the velocity deviation from circular motion is larger after crossing the dust lanes and there is gas streaming motion along the dust lanes as predicted by hydrodynamic models \citep{Athanassoula92}.
The black dashed and dotted lines in panel (c) and (d) show the identified nuclear ring and spiral structures as presented in Fig.~\ref{fig:channel_maps_zoomin}. The non-zero residual velocity along the spiral structures may indicate gas inflows toward the galactic central region.

\section{Streaming Motion}
\begin{center}
\begin{figure*}
  \includegraphics[scale=0.7, clip=true, trim = 0 0 0 0]{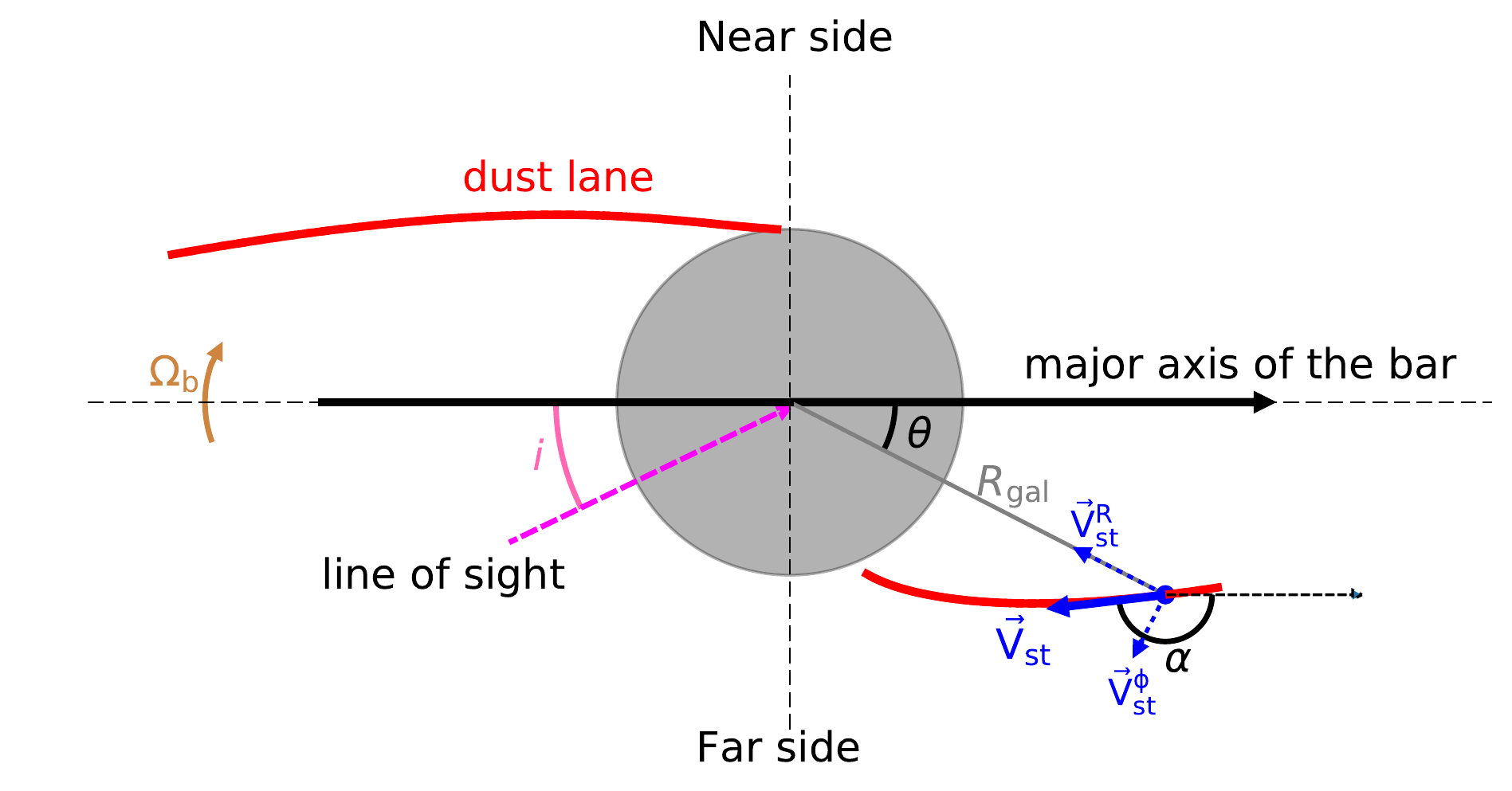}
 \caption{Schematic diagram of the dust lanes and the streaming velocity $\vec{\rm V}_{\rm st}$ on the galactic plane. $\theta$ is the angle between the major axis of the bar and the location of a point in the dust lane. $\alpha$ is the angle between the major axis of the bar and the streaming velocity $\vec{\rm V}_{\rm st}$.  $\vec{\rm V}_{\rm st}^{\rm R}$ and $\vec{\rm V}_{\rm st}^{\rm \phi}$ are $\vec{\rm V}_{\rm st}$ projected to the radial and azimuthal directions, respectively. 
 $\Omega_{\rm b}$ represents the pattern speed of the bar and $i$ is the inclination of the galaxy.
Note that the major axis of the bar is identical to the major axis of the galaxy in this diagram, such as in NGC\,3504 \citep{kuno00}.
 }
\label {fig:schematic}
\end{figure*}
\end{center}

We calculate the gas streaming velocity along the dust lanes assuming that (1) the gas moves along the dust lanes in the rotating frame of the bar, and (2) the circular motions are well described by the rotation curve derived from {\it Kinemetry}.
The first assumption is predicted by hydrodynamic models and revealed in observations, especially in explaining the velocity gradient across dust lanes and the negative/positive residual velocity on the far/near side of the dust lanes \citep{Athanassoula92,Benedict1996, Regan97,regan99,Schinnerer2002,Perez2004,Perez2008, Fanali2015}. It also implies that the velocity of the gas in the dust lanes on the galactic plane consists of the rotational velocity of the bar $R_{\rm gal}\Omega_{\rm b}$ and the gas streaming velocity ${\rm V}_{\rm st}$ along the dust lanes.
The streaming velocity ${\rm V}_{\rm st}$ can be decomposed into the radial and azimuthal directions, ${\rm V}_{\rm st}^{\rm R}$ and ${\rm V}_{\rm st}^{\rm \phi}$, as shown in the scheme in Fig.~\ref{fig:schematic}. Therefore, the observed velocity in the dust lanes is 
\begin{equation}
{\rm V}_{\rm obs}={\rm V}_{\rm sys} + \left( {\rm V}_{\rm st}^{\rm \phi} + {R_{\rm gal} \Omega_{\rm b}} \right) \sin(\frac{\pi}{2}+\theta) \sin i + {\rm V}_{\rm st}^{\rm R} \sin(\pi+\theta) \sin i,
\label {eq:V_obs}
\end{equation}
where ${\rm V}_{\rm sys}$ is the velocity of the galaxy and $\theta$ is the angle between the major axis of the bar and the position ${R_{\rm gal}}$. Note that the second and third terms represent the azimuthal and radial components, respectively. In the case of NGC\,3504, because the PA of the bar is similar to the PA of the galaxy \citep{kuno00} (i.e. the major axis of the bar and that of the galaxy are similar), we assume that they are identical in our calculation. ${\rm V}_{\rm st}^{\rm R}$ and ${\rm V}_{\rm st}^{\rm \phi}$ in Equation (\ref{eq:V_obs}) can be expressed as
\begin{equation}
{\rm V}_{\rm st}^{\rm R}={\rm V}_{\rm st}\cos(\pi-\alpha+\theta) = -{\rm V}_{\rm st}\cos(\alpha-\theta),
\label {eq:V_st_R}
\end{equation}
\begin{equation}
{\rm V}_{\rm st}^{\rm \phi}={\rm V}_{\rm st}\cos(\alpha-\frac{\pi}{2}-\theta)={\rm V}_{\rm st}\sin(\alpha-\theta),
\label {eq:V_st_phi}
\end{equation}
where $\alpha$ is the angle between the streaming velocity ${\rm V}_{\rm st}$ and the major axis of the bar.

Furthermore, under the second assumption, the velocity component in the azimuthal direction can be subtracted from the velocity field, so that the residual velocity ${\rm V}_{\rm res}$ in the dust lanes in Fig.~\ref{fig:model_res_maps} (b) only consists of the third term on the right-hand side of Equation (\ref{eq:V_obs}):
\begin{equation}
{\rm V}_{\rm res} = {\rm V}_{\rm st}^{\rm R} \sin(\pi+\theta) \sin i.
\label {eq:res}
\end{equation}
Thus, combining Equation (\ref{eq:V_st_R}) and (\ref{eq:res}), we obtain the streaming velocity 
\begin{equation}
{\rm V}_{\rm st} = \frac{{\rm V}_{\rm res}}{\cos(\alpha-\theta)\sin \theta \sin i}~.
\label {eq:streamV}
\end{equation}
Note that $\theta$ and $\alpha$ can be calculated by $\theta=\tan^{-1}\left(\frac{\tan \theta^\prime}{\cos i } \right)$ and $\alpha=\tan^{-1}\left(\frac{\tan \alpha^\prime }{\cos i} \right)$, where $\theta^\prime$ and $\alpha^\prime$ are defined in the same way as $\theta$ and $\alpha$ but in the plane of the sky. 

\begin{figure*}
\includegraphics[scale=0.29, clip=true, trim = 0 0 0 0]{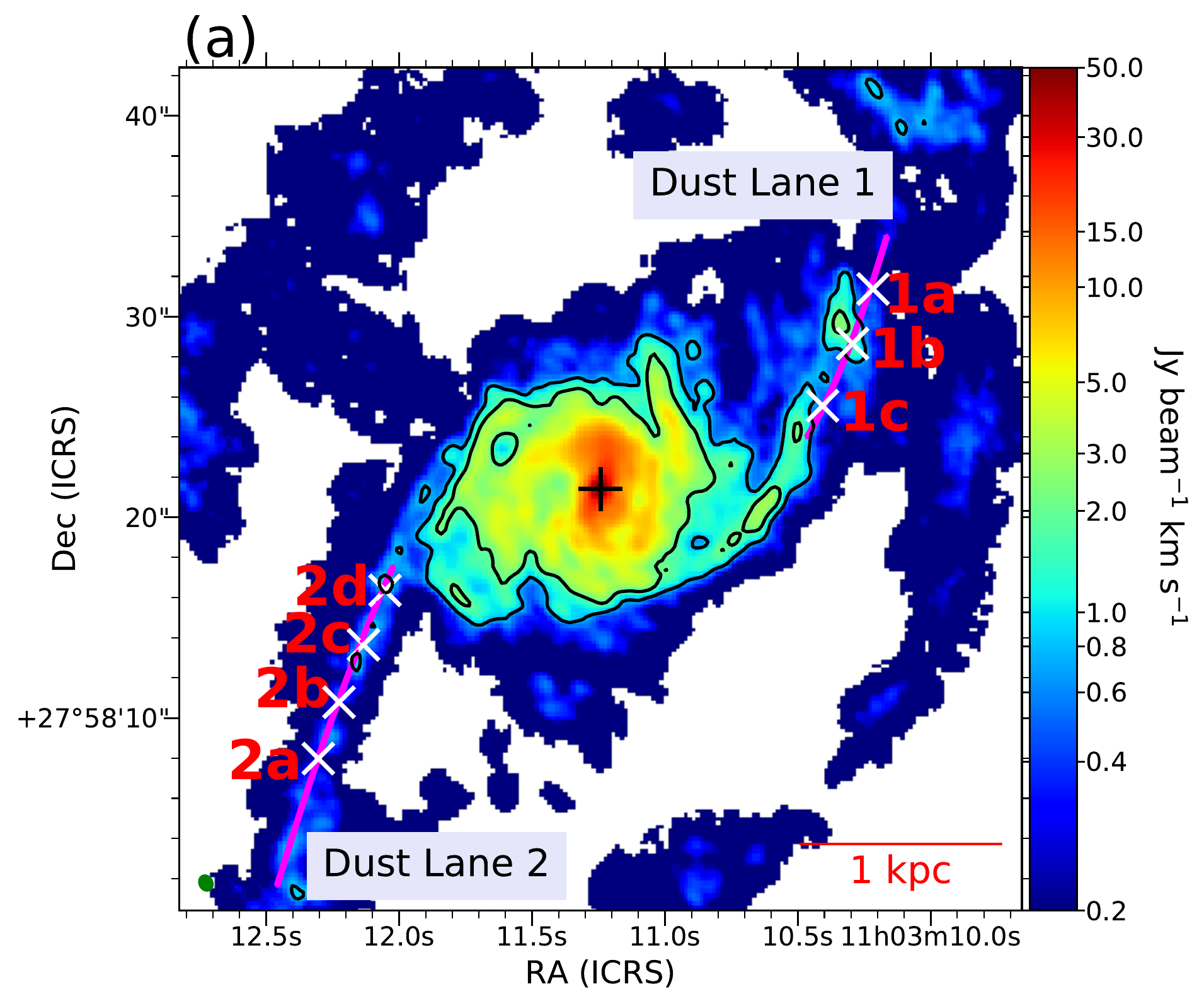}
  \includegraphics[scale=0.47, clip=true, trim = 0 0 5 0]{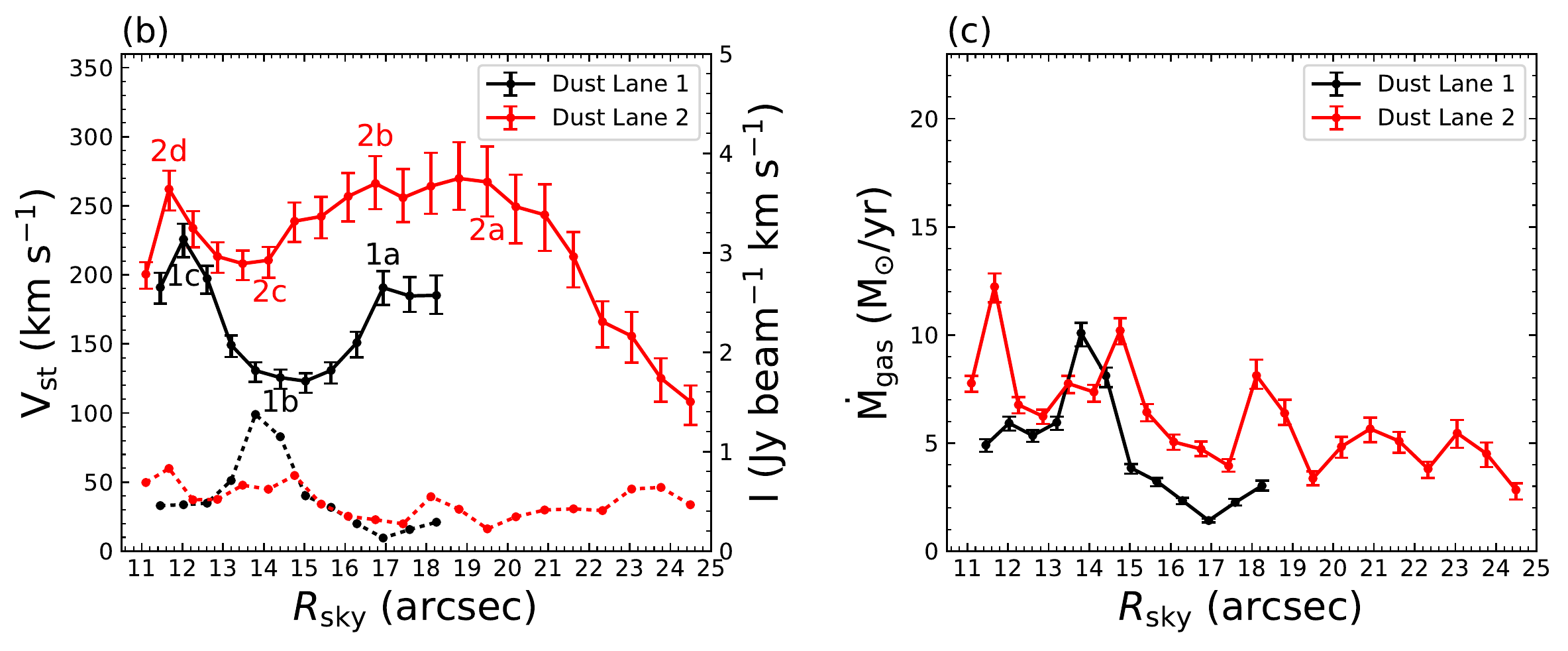}
 \caption{(a) Illustration of the gas streaming along the two dust lanes, marked as Dust Lane 1 and Dust Lane 2, as indicated by the magenta lines superimposed on the ALMA CO (2--1) integrated intensity (moment 0) map. The black contours of the CO (2--1) integrated intensity are at levels of 0.8 and 2.0\,Jy\,beam$^{-1}$. The letters 1a to 2d indicate the locations corresponding to the radii given in panel (b). 
 (b) Gas streaming velocity (solid lines) and CO (2--1) integrated intensity (dashed lines) along the dust lanes as a function of radius $R_{\rm sky}$. (c) Same as panel (b), but for the gas mass inflow rate, $\dot{\rm M}_{\rm gas}$.}
\label{fig:Vst_R}
\end{figure*}

Using Equation (\ref{eq:streamV}), we can calculate the gas streaming velocity along the dust lanes. The location of the dust lanes are identified on the SDSS $g$-band image (Fig.~\ref{fig:mom0_map} (b)). Fig~\ref{fig:Vst_R} (a) shows the dust lanes, marked as Dust Lane 1 and 2, superimposed on the ALMA CO (2--1) integrated intensity (moment 0) map.

Fig~\ref{fig:Vst_R} (b) shows the gas streaming velocity (solid lines) and the integrated intensity (dashed lines) along the dust lanes and toward the galactic center region to $R_{\rm sky}\approx 11\arcsec$. 
The average streaming velocities on Dust Lane 1 and 2 are about 165 and 221\kms, respectively. 
The letters 1a to 2d indicate the radii where the gas streaming velocity curves have a local minimum or change significantly. Their corresponding locations are also marked on the integrated intensity map, as shown in Fig.~\ref{fig:Vst_R} (a). 
 Comparing the information of the gas streaming velocities and the CO (2--1) integrated intensities in Fig~\ref{fig:Vst_R} (b), we find that the gas streaming velocities decrease when the gas intensities increase in Dust Lane 1. The gas streaming velocity in Dust Lane 1 starts to drop significantly at point 1a when the integrated intensity increases from 0.13\,Jy\,beam$^{-1}$\kms\ at point 1a to 1.15\,Jy\,beam$^{-1}$\kms\ at point 1b. Then, the streaming velocity increases toward point 1c when the integrated intensity decreases.
The same phenomenon also happens to a smaller extent in  Dust Lane 2. At point 2a, the increase of the streaming velocity slows down and the streaming velocity keeps being relatively constant toward point 2b. Then, the gas streaming velocity starts to decrease from point 2b to point 2c, where the integrated intensity is higher. Finally, the streaming velocity increases toward point 2d where the integrated intensity decreases.
At the region where the radius is smaller than that of point 1c and 2d, the motion of the gas might be affected by the inner molecular region and is not considered here.

Fig.~\ref{fig:Vst_R} (c) shows the mass inflow rate in the dust lanes.
Using Equation (\ref{eq:gas_mass}) to derive the gas mass and assuming the width of the streaming gas lane is 1\farcs5, the mass inflow rate ($\dot{\rm M}_{\rm gas}$) at the position R in the dust lane can be calculated as in
 \citet{Regan97}:
 \begin{equation}
  \dot{\rm M}_{\rm gas}({\rm R}) = \Sigma_{\rm gas}({\rm R}) {\rm W} {\rm V}_{\rm st}({\rm R}),
\label {eq:mass_flux}
\end{equation}
where $\Sigma_{\rm gas}$ is the gas mass surface density, W is the width of the streaming gas lane and ${\rm V}_{\rm st}$ is the gas streaming velocity as calculating using Equation (\ref{eq:streamV}). From Fig.~\ref{fig:Vst_R} (c), it can be seen that the average of the gas  inflow rate  along each of the two dust lanes is about 6\,M$_\odot$ yr$^{-1}$ with a slight decline towards larger radius. Thus, the total inflow rate along two dust lanes is 12\,M$_\odot$ yr$^{-1}$.
This estimated mass inflow rate is comparable to the typical mass inflow rate of the order of 0.1--10\,M$_\odot$\,yr$^{-1}$  found in previous works for galaxies, such as NGC\,1530, NGC\,3368, NGC\,4736 and NGC\,5248 \citep{Regan97,Sakamoto99, Haan09}. The declining trend of the mass inflow rate could be due to the lower gas streaming velocity and/or the lower gas mass surface density at larger radius. This phenomenon is also found in the western dust lane in NGC\,1530 \citep{Regan97}. 

\section{The Pattern Speed of the Bar}
Under the same assumptions described in Section 5, the velocity in the dust lanes on the modelled velocity field in Fig.~\ref{fig:model_res_maps} (a) consists of the first and second terms of the right-hand side of Equation (\ref{eq:V_obs}):
\begin{equation}
{\rm V}_{\rm mod}={\rm V}_{\rm sys}+\left( {\rm V}_{\rm st}^{\rm \phi} + {R_{\rm gal}\Omega_{\rm b}} \right) \cos \theta \sin i,
\label {eq:V_model}
\end{equation}
where $R_{\rm gal}= R_{\rm sky} \sqrt{\cos^2 {\theta^{\prime}} + \sin^2 \theta^{\prime}/ \cos^2 i}$.
Therefore, combining Equation (\ref{eq:V_st_phi}) and  (\ref{eq:V_model}), we can get the pattern speed of the bar:
\begin{equation}
\Omega_{\rm b} = \frac{{\rm V}_{\rm mod} - {\rm V}_{\rm sys} - {\rm V}_{\rm st}\sin(\alpha-\theta) \cos \theta \sin i}{{R_{\rm gal}} \cos \theta \sin i}~.
\label {eq:Omega_bar}
\end{equation}

\begin{figure*}
\includegraphics[scale=0.5, clip=true, trim = 0 0 0 0]{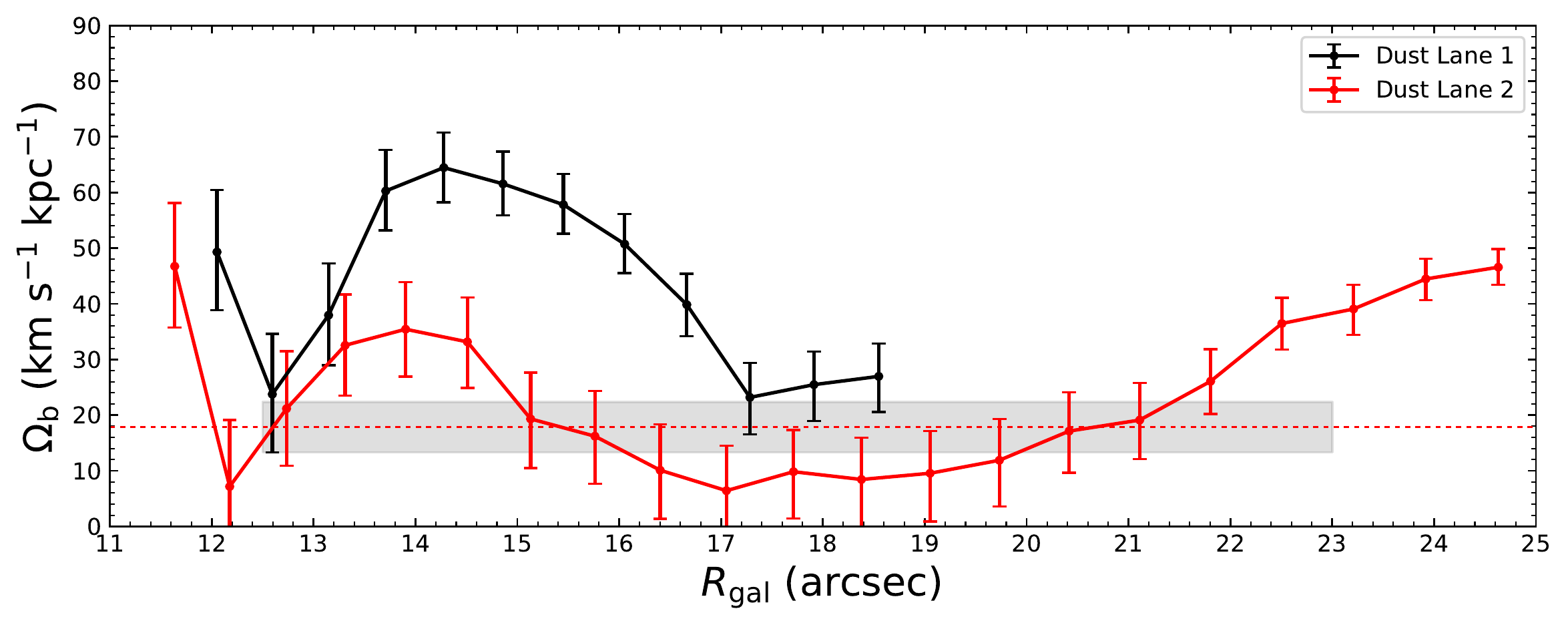}
\caption{$\Omega_{\rm b}$ along Dust Lane 1 and 2 as function of the radius on the galactic plane, as shown by the black and red solid lines, respectively. The red dashed line represents the derived pattern speed of the bar, $\Omega_{\rm Bar}=18\pm5$\kmskpc, which is the least-squares fit to the data points at $12\farcs5<R_{\rm gal}<23\arcsec$. The grey area is the 95 per cent confidence interval.}
\label{fig:Omega_bar}
\end{figure*}

Using the systemic velocity ${\rm V}_{\rm sys}=1532.2$\kms\ and the derived ${\rm V}_{\rm st}$ in  Dust Lane 1 and 2 as mentioned in Section 5, Fig.~\ref{fig:Omega_bar} shows the estimated pattern speed of the bar at different radius $R_{\rm gal}$ on the galactic plane.
The errors indicate the variances caused by the uncertainty of the inclination and PA.
At $R_{\rm gal}<12\farcs5$, the estimated pattern speed has lower confidence since the motion of the gas might be affected by the inner molecular region, therefore not satisfying our first assumption in Section 5, that is the gas moves along the dust lanes in the rotating frame of the bar. 
Furthermore, in Dust Lane 1, the existence of large gas clumps could change the motion of the gas and therefore render the first assumption invalid and may cause a larger variation of the streaming velocity and $\Omega_{\rm b}$.
On the other hand, we note that the integrated intensity is nearly the same along Dust Lane 2, and the pattern speed is relatively constant at $12\farcs5<R_{\rm gal}<23\arcsec$ in Dust Lane 2. 
Therefore, we performed least-squares fitting of the data points at $12\farcs5<R_{\rm gal}<23\arcsec$ of the Dust Lane 2 to obtain the pattern speed of the bar, and obtained $\Omega_{\rm Bar}=18\pm5$\kmskpc. 

Although $\Omega_{\rm Bar}$ is only derived from Dust Lane 2, from Fig~\ref{fig:Omega_bar}, we can see that at $17\arcsec<R_{\rm gal}<19\arcsec$ in Dust Lane 1, where there are no large clumps and the motion of gas is assumed to satisfy the first assumption in Section 5, the estimated pattern speeds are also consistent with our derived pattern speed $\Omega_{\rm Bar}$. This implies that our derived pattern speed $\Omega_{\rm Bar}$ could be valid for both sides of the bar, even when it is derived using only the Dust Lane 2.

With the derived pattern speed of the bar $\Omega_{\rm Bar}=18$\kmskpc\ and the frequency curve $\Omega-\kappa/2$, we found that the inner inner Lindblad resonance (IILR) is located at $R\sim 0.2$\,kpc, as shown in Fig.~\ref{fig:model_rot_curve}. 
The outer inner Lindblad resonance (OILR) is predicted to be between 1.9 and 2.3\,kpc due to the uncertainty of $\Omega-\kappa/2$ in this region.
The epicyclic frequency $\kappa$ was calculated from 
\begin{equation}
\kappa = \sqrt{4\Omega^2+R\frac{d\Omega^2}{dR}}.
\label {eq:kappa}
\end{equation}

\cite{kenney93} estimated the pattern speed of the bar in NGC\,3504 to be $\sim 77$\kmskpc, assuming the corotation radius is located at the end of the bar. 
 On the other hand, \cite{kuno00} obtained the upper limit of the pattern speed of the bar, $\sim 41$\kmskpc, based on the two assumptions that the gas flows along the dust lane toward the galactic center and that the 
 misalignment between the major axis of the bar and the major axis of the galaxy is small. Furthermore, they suggested that the corotation radius is at least two times larger than the radius of the bar if that upper limit is close to the pattern speed.
 Our result of $\Omega_{\rm Bar}=18\pm5$\kmskpc\ is consistent with the results of \cite{kuno00} and has a smaller value due to the consideration of the gas streaming velocity $\rm V_{st}$ in Equation (\ref{eq:Omega_bar}).

\section{Discussion}
\subsection{The Nuclear Ring}
Regarding the identified nuclear ring as shown in Fig.~\ref{fig:channel_maps_zoomin}, we applied  least-squares fitting to an ellipse to determine the center of the ring at ($11^{\rm{h}} 03^{\rm{m}} 11\fs29$, $+27\degr58 \arcmin 21\farcs31$) with semi-major and semi-minor axes of about $3\farcs6$ and $3\farcs2$, and a PA of $-44\fdg4$. 
The derived ellipticity of the ring is 0.11, which is in the range of 0--0.4 reported by \cite{Comeron10}, based on a sample of more than 100 galaxies. 
In addition, the difference between the center of the ellipse and the center of the galaxy is about $0\farcs7$, which is comparable to the size of our synthesized beam. Therefore, high resolution data is needed to confirm whether this discrepancy is real or not.  

Furthermore, with the semi-major and semi-minor axes of the ring of $\sim3\farcs6$ and $3\farcs2$, it seems that the inner ring is not located at either IILR or OILR, but between the two ILRs and close to the IILR, as can be seen from Fig.~\ref{fig:model_rot_curve}. Inner rings located between the two ILRs are also found in other galaxies, such as NGC 1097, IC 1438, IC 4214, NGC 1512 and NGC 6753 \citep{Pinol14, Schmidt19} and could be explained by different scenarios. For example, \cite{Combes96} proposed that a nuclear ring forms between the IILR and the OILR when there are two ILRs. \cite{Sormani18} suggested that the gas follows the $x_2$ orbits and forms a ring in the reversed shear region (see their Fig.~3).
However, it is still hard to predict the location and the size of the nuclear ring as it depends on several factors, such as the random turbulent 
motion of interstellar clouds, the sound speed of the gas, the pattern speed of the bar, and the bar strength \citep{Patsis00, Kim12, Seo19, Schmidt19}. Nevertheless, our results confirm the existence of the nuclear molecular ring between two ILRs and close to the IILR.

\subsection{The Inner Gas Bar}
As mentioned in Section 1, to understand whether the gas is involved in the formation of the inner bar, the first critical step is to investigate the morphology and the mass of the molecular gas in double-barred galaxies. \cite{Petitpas02} studied two candidate double-barred galaxies and found that only NGC\,2273 has an inner gas bar, although both NGC\,2273 and NGC\,5728 have large amounts of molecular gas in their nuclei.
NGC\,3504, which was classified as a candidate double-barred galaxy \citep{Perez00}, presents an inner gas bar as can be seen in Fig.~\ref{fig:mom0_map_zoomin} (a). 
The position angle and the length of the inner gas bar are about the same as the inner stellar bar, as shown in Fig.~\ref{fig:mom0_map_zoomin} (f).
From the CO velocity field as shown in Fig.~\ref{fig:mom0_map_zoomin}(h), the twisted and tilted isovelocity contours relative to the minor axis of the galaxy at the central region $R<2\farcs5$ strongly suggest the presence of an inner bar in NGC\,3504. This is also supported by the indication of non-circular motion in the PV diagrams in Fig.~\ref{fig:PVD_MajnMin} and the clear evidence for radial gas motions in the central 3\arcsec\ shown in Fig.~\ref{fig:model_coefficient}.

Furthermore, given an stellar mass of $1.84 \times 10^{10}$ M$_{\odot}$ \citep{Font17} and a molecular gas mass $\sim 3.1 \times 10^9$ M$_{\odot}$, as mentioned in Section 3.2.2, the gas mass in NGC\,3504 is about 17 per cent of the stellar mass. 
This agrees with the simulation result in \cite{Friedli93} where the required gas mass to form a double-barred galaxy is at least $\sim$ 10 per cent of the stellar mass of the galaxy. 
Therefore, our result of NGC\,3504 gives a new example of an inner gas bar within a gas-rich double-barred galaxy, and supports the scenario where the formation of double-barred galaxies could be associated with the existence of molecular gas, as mentioned in Section 1.

\subsection{Gas Flow in the Bar}
Considering the average streaming velocity of 165 and 221\kms\ along Dust Lane 1 and 2 (see Section 5) and the deprojected bar radius of 32\arcsec\ (3.2\,kpc) \citep{kenney93}, the time for the gas to fall in from the end of the bar to the inner molecular region at $R_{sky}<11\arcsec$ (1.1\,kpc) along Dust Lane 1 and 2 are $\sim12$ and 9\,Myr, respectively. Note that the these estimated times could be a factor of two longer because the streaming velocity becomes smaller when the local gas density increases.

At the central region $R<10$\arcsec, since the molecular gas mass is 75 per cent of the total molecular gas mass ($ 3.1 \times 10^9$\,M$_{\odot}$) (see Section 3.2)
and the star formation rate estimated from the Lyman continuum photon rate in the same region is 2.3\,M$_{\odot}\,{\rm yr}^{-1}$ \citep{kenney93}, the average gas consumption time would be about 1\,Gyr. 
Furthermore, if we consider the additional 25 per cent of the total gas mass at $R>10$\arcsec\ and assume it will accrete to the central region at $R<10$\arcsec\ at the total inflow rate of 12\,M$_{\odot}\,{\rm yr}^{-1}$ (See Section 5), this gas will take an additional $\sim$ 64\,Myr to reach the central region. 
At the star formation rate of 2.3\,M$_{\odot}\,{\rm yr}^{-1}$, the gas consumption time for all total molecular gas would be about 1.3\,Gyr.

\section{Conclusions}
We presented the morphology and kinematics of the CO (2--1) line in the galaxy NGC\,3504, as observed by ALMA. The achieved linear resolution of less than 80\,pc allowed us to
reveal the non-axisymmetric symmetric structures in the central region and also to investigate the kinematics of the molecular gas. Our main results are:\\
\begin{enumerate}[align=left,label=(\roman*), nosep]
\item Both axisymmetric and non-axisymmetric structures, including the inner molecular gas bar, the nuclear ring, 
 and the nuclear spirals, are found in the central 1\,kpc region. These inner structures were not recognized in previous publications \citep{kenney93, kuno00} due to the lack of sufficient angular resolution.\\
\item The estimated total molecular mass is about $3.1\times 10^9\,{\rm M}_{\odot}$, corresponding to 17 per cent of the stellar mass.\\
\item {\it Kinemetry} was employed to fit the velocity field of NGC\,3504. The fit results are the systemic velocity of the galaxy of $1532.2 \pm 0.2$\kms, the inclination of $25\degr \pm 1\degr$, and PA=$153\degr \pm 2\degr$, which are consistent with the values reported in the literature \citep{grosbol85, kenney93}.\\
\item Circular motion strongly dominates at $R= 3\arcsec-8$\arcsec\ ($\sim 0.3 - 0.8$\,kpc), but radial motion becomes important in the regions where the bars are present, corresponding to $R <$ 3\arcsec\  (0.3\,kpc) and $R=  10\arcsec-25\arcsec\ (1.0 - 2.5$\,kpc).\\

\item The gas streaming velocities along the two dust lanes were derived from the velocity residual map by assuming that the gas moves along the dust lanes in the bar rotating frame. 
The velocity residual map was obtained by subtracting the fitted circular motion from the velocity field.
We found that the gas streaming velocity on the dust lane varies and presents an anti-correlation with the gas surface density. 
The average streaming velocities on each of the two dust lanes are about 165 and 221\kms, and the total gas inflow rate along the two dust lanes is about $\dot{\rm M}_{\rm gas}$ = 12\,{\rm M}$_{\odot}$\,yr$^{-1}$.\\
\item Following the same assumption as above and also using the derived streaming velocities on the dust lanes, we calculate the pattern speed of the bar to be $18\pm5$\kmskpc. This result is consistent with the upper limit of 41\kmskpc\ derived by \citet{kuno00}.\\
\end{enumerate}

In simulation studies, the inner bar in double-barred galaxies could be formed either with or without gas \citep{Friedli93, Shen09, Du15}. The main differences between these two scenarios are the presence of the inner gas bar and the amount of gas mass of at least 10 per cent of the stellar mass of the galaxy.
Therefore, to distinguish between these two scenarios, a major step is to resolve the inner gas bar and to determine the gas mass in double-barred galaxies. 
Based on our results, we conclude that the existence of the inner gas bar and the large amount of gas in NGC\,3504 support the scenario for the formation of double-barred galaxies associated with the existence of molecular gas.

\section*{Acknowledgements}
This paper makes use of the following ALMA data: ADS/JAO.ALMA\#2016.1.00650.S. ALMA is a partnership of ESO (representing its member states), NSF (USA) and NINS (Japan), together with NRC (Canada), MOST and ASIAA (Taiwan), and KASI (Republic of Korea), in cooperation with the Republic of Chile. The Joint ALMA Observatory is operated by ESO, AUI/NRAO and NAOJ. The authors are grateful to \citeauthor*{kraj06} for the Kinemetry code. The authors thank the referee, Prof. Jeffrey Kenney, for valuable comments and suggestions that improved this paper. The authors also thank the support of the Astronomical Data Center of the National Astronomical Observatory of Japan, and Anthony Moraghan for assistance on the computational facilities and resources at Academia Sinica Institute of Astronomy and Astrophysics (ASIAA).
YTW wishes to thank Dr. Yu-Nung Su and Dr. Sheng-Yuan Liu for hosting her visit at ASIAA in 2020.

\section*{Data availability}
The ALMA data underlying this article is available online in the ALMA archive at https://almascience.nao.ac.jp/aq/, with project code \#2016.1.00650.S. This study has also used SDSS data provided at https://dr12.sdss.org/fields.

\bibliographystyle{mnras}
\bibliography{Ting_ref}

\bsp	
\label{lastpage}
\end{document}